\documentclass[12pt]{article}

\title{Kinetic and hydrodynamic models of chemotactic aggregation}

\usepackage{amsthm,amsmath,amssymb,a4wide,psfig,epsf}
\def\mb#1{\setbox0=\hbox{$#1$}\kern-.025em\copy0\kern-\wd0
\kern-0.05em\copy0\kern-\wd0\kern-.025em\raise.0233em\box0}

\begin{document}

\author{Pierre-Henri Chavanis and Cl\'ement Sire}
\maketitle
\begin{center}
Laboratoire de Physique Th\'eorique (IRSAMC, CNRS), Universit\'e Paul
Sabatier,\\ 118, route de Narbonne, 31062 Toulouse Cedex, France\\
E-mail: {\it chavanis{@}irsamc.ups-tlse.fr \&
clement.sire{@}irsamc.ups-tlse.fr }

\vspace{0.5cm}
\end{center}

\begin{abstract}

We derive general kinetic and hydrodynamic models of chemotactic
aggregation that describe certain features of the morphogenesis of
biological colonies (like bacteria, amoebae, endothelial cells or
social insects). Starting from a stochastic model defined in terms of
$N$ coupled Langevin equations, we derive a nonlinear mean field
Fokker-Planck equation governing the evolution of the distribution
function of the system in phase space.  By taking the successive
moments of this kinetic equation and using a local thermodynamic
equilibrium condition, we derive a set of hydrodynamic equations
involving a damping term. In the limit of small frictions, we obtain a
hyperbolic model describing the formation of network patterns
(filaments) and in the limit of strong frictions we obtain a parabolic
model which is a generalization of the standard Keller-Segel model
describing the formation of clusters (clumps). Our approach connects
and generalizes several models introduced in the chemotactic
literature. We discuss the analogy between bacterial colonies and
self-gravitating systems and between the chemotactic collapse and the
gravitational collapse (Jeans instability).  We also show that the
basic equations of chemotaxis are similar to nonlinear mean field
Fokker-Planck equations so that a notion of effective generalized
thermodynamics can be developed.

\vskip1cm


\end{abstract}

\eject

\section{Introduction}
\label{intro}

In many fields of physical sciences, one is confronted with the
description of the evolution of a system of particles which
self-consistently attract each other over large distances
\cite{houches,cras}. This is the case in biology, for example, in
relation with the process of chemotaxis \cite{murray}. Chemotaxis
explains the spontaneous self-organization of biological cells
(bacteria, amoebae, endothelial cells,...) or even insects (like
ants) due to the long-range attraction of a chemical (pheromone,
smell, food,...) produced by the organisms themselves. The
chemotactic aggregation of biological populations is usually studied
in terms of the Keller-Segel model consisting in two coupled
differential equations \cite{ks}:
\begin{equation}
{\partial\rho\over\partial t}=D_*\Delta\rho-\chi\nabla\cdot
(\rho\nabla c), \label{intro1}
\end{equation}
\begin{equation}
\label{intro2}{\partial c\over\partial t}=-k(c)c+h(c)\rho+D_{c}\Delta
c,
\end{equation}
which describe the evolution of the concentration $\rho({\bf r},t)$ of
the biological organisms and of the secreted chemical $c({\bf
r},t)$. The Keller-Segel model is a {\it parabolic model} where the
evolution of the concentration of the biological organisms is governed
by a drift-diffusion equation (\ref{intro1}). The diffusion models the
erratic motion of the particles (like in Brownian theory) and the
drift term models a systematic motion along
the gradient of concentration of the secreted chemical. When $\chi>0$
the cells are attracted in regions of high concentration while for
$\chi<0$ they are repelled from the regions of high concentration (in
that case the chemical acts as a poison). The
evolution of the secreted chemical is described by a diffusion
equation (\ref{intro2}) involving terms of source and degradation: the
chemical is produced by the organisms at a rate $h(c)$ and it is
degraded at a rate $k(c)$. For $\chi>0$, the Keller-Segel model is
able to reproduce the chemotactic aggregation (collapse) of biological
populations when the attractive drift term $\chi \rho \|\nabla c\|$
overcomes the diffusive term $D_{*}\|\nabla
\rho\|$ above a critical mass $M_{c}$ \cite{horstmann}. This is similar to the
gravitational collapse of self-gravitating Brownian particles,
described by the Smoluchowski-Poisson system, below a critical
temperature $T_{c}$ \cite{crs} (see \cite{crrs,mass} for a detailed
discussion of the analogy between the Keller-Segel model and the
Smoluchowski-Poisson system). These parabolic models ultimately lead
to the formation of Dirac peaks \cite{sc,post}.

Some regularizations of the Keller-Segel model have been
introduced. They have the form of generalized drift-diffusion
equations \cite{gen,crrs}:
\begin{equation}
{\partial\rho\over\partial t}=\nabla \cdot \left (\chi \left (\nabla p-\rho\nabla c\right )\right ), \label{add1}
\end{equation}
\begin{equation}
\label{add2}{\partial c\over\partial t}=-k(c)c+h(c)\rho+D_{c}\Delta
c,
\end{equation}
in which the simple diffusion term $D_{*}\nabla \rho$ in
Eq. (\ref{intro1}) is replaced by a more general ``pressure term''
$\nabla p(\rho)$. The drift-diffusion equation (\ref{add1}) is similar
to a generalized Smoluchowski equation \cite{next,cban} \footnote{The
possibility that the diffusion coefficient $D_{*}$ and the chemotactic
sensitivity $\chi$ depend on the density of cells $\rho$ and on the
density of chemical $c$ is considered in the primitive model of Keller
\& Segel \cite{ks} (see Appendix \ref{sec_prim}).  The very much 
studied model (\ref{intro1})-(\ref{intro2}) is a simplification of the
primitive Keller-Segel model where the coefficients $D_{*}$ and $\chi$
are assumed constant.}. By adapting the barotropic equation of state
$p(\rho)$, one can obtain regularized chemotactic models preventing
the density from reaching infinitely large values and forming
singularities \cite{degrad}. In that case, the Dirac peaks (clumps)
are replaced by smoother density profiles (aggregates). The dynamical
evolution of the regularized model (\ref{add1})-(\ref{add2})
generically leads to the formation of ${\cal N}(t)$ round aggregates
which progressively merge until only one big aggregate remains at the
end.

However, recent experiments of {\it in vitro} formation of blood
vessels show that cells randomly spread on a gel matrix autonomously
organize to form a connected vascular network that is interpreted as
the beginning of a vasculature \cite{gamba}. This phenomenon is
responsible of angiogenesis, a major actor for the growth of
tumors. These networks cannot be explained by the parabolic models
(\ref{intro1})-(\ref{add2}) that lead to pointwise blow-up or
round aggregates. However, they can be recovered by {\it
hyperbolic models} that lead to the formation of networks patterns
that are in good agreement with experimental results. These models
take into account inertial effects and they have the form of
hydrodynamic equations \cite{gamba}:
\begin{equation}
\label{intro3} {\partial\rho\over\partial t}+\nabla\cdot (\rho{\bf u})=0,
\end{equation}
\begin{equation}
\label{intro4} \frac{\partial {\bf u}}{\partial t}+({\bf u}\cdot
\nabla){\bf u}=-{1\over\rho}\nabla p+\nabla c,
\end{equation}
\begin{equation}
\label{intro5}{\partial c\over\partial t}=-k(c)c+h(c)\rho+D_{c}\Delta c.
\end{equation}
The inertial term models cells directional persistence and the general
density dependent pressure term $-\nabla p(\rho)$ can take into
account the fact that the cells do not interpenetrate. In these
models, the particles concentrate on lines or filaments. These
structures share some analogies with the formation of ants' networks
(due to the attraction of a pheromonal substance) and with the
large-scale structures in the universe that are described by similar
hydrodynamic (hyperbolic) equations: the so-called Euler-Poisson
system. The similarities between the networks observed in astrophysics
(see Figs 10-11 of \cite{vergassola}) and biology (see Figs 1-2 of
\cite{gamba}) are striking.

The above-mentioned parabolic and hyperbolic models are continuous
models which describe the evolution of a smooth density field
$\rho({\bf r},t)$ and, in the case of hyperbolic models, a smooth
velocity field ${\bf u}({\bf r},t)$. In this paper, we propose a
kinetic derivation of these models starting from a microscopic
description of the dynamics of the biological population. We introduce
stochastic equations for the motion of each individual and,
implementing a mean field approximation, we obtain the corresponding
Fokker-Planck equation governing the evolution of the distribution
function $f({\bf r},{\bf v},t)$ of the system in phase space. The
stochastic Langevin equations involve a friction force, an effective
force due to the chemotactic attraction of the chemical and a random
force (noise) whose strength can depend on the local distribution
function itself. This dependence can take into account microscopic
constraints (``hidden constraints'') that affect the dynamics of the
particles at small scales. We can have (i) {\it close packing
effects} (like finite size effects, excluded volume constraints,
steric hindrance...) that forbid the interpenetration of the particles
and prevent the system from reaching arbitrarily high densities
\cite{degrad} and (ii) {\it nonextensivity effects} that alter the usual
random walk and lead to anomalous diffusion and non-ergodic behaviour
\cite{lang,logotropes}. The resulting generalized stochastic equations
lead to nonlinear mean field Fokker-Planck equations similar to those
occurring in the context of generalized thermodynamics
\cite{frank,next,gen}. Therefore, as first noticed in
\cite{gen}, the chemotaxis of biological populations can be a physical
system where a notion of effective generalized thermodynamics
applies. By taking the successive moments of these generalized
Fokker-Planck equations and using a local thermodynamic equilibrium
condition, we derive a closed set of hydrodynamic equations
\begin{equation}
\label{intro3ht} {\partial\rho\over\partial t}+\nabla\cdot (\rho{\bf u})=0,
\end{equation}
\begin{equation}
\label{intro4ht} \frac{\partial {\bf u}}{\partial t}+({\bf u}\cdot
\nabla){\bf u}=-{1\over\rho}\nabla p+\nabla c-\xi\rho {\bf u},
\end{equation}
\begin{equation}
\label{intro5ht}{\partial c\over\partial t}=-k(c)c+h(c)\rho+D_{c}\Delta c,
\end{equation}
involving a friction term $-\xi {\bf u}$ \cite{gen,epjbjeans}.  For
$\xi=0$, we recover the hyperbolic model (\ref{intro3})-(\ref{intro5})
proposed by Gamba {\it et al.}
\cite{gamba} to model vasculogenesis and for $\xi\rightarrow
+\infty$ we obtain the parabolic model (\ref{add1})-(\ref{add2})
generalizing the Keller-Segel model (\ref{intro1})-(\ref{intro2}).  We
discuss the analogy between bacterial colonies and self-gravitating
systems and between the chemotactic collapse and the ``gravitational
collapse'' (Jeans instability) \cite{crrs,mass,epjbjeans}. Indeed, the
kinetic and hydrodynamic models of biological populations derived in
this paper are similar to those describing self-gravitating systems
\cite{gen,virial}. Therefore, our approach connects various topics studied
by different communities: systems with long-range interactions
\cite{houches}, nonlinear mean field Fokker-Planck equations and
generalized thermodynamics
\cite{frank}, self-gravitating systems \cite{revue} and chemotaxis
\cite{ks}.

\section{A stochastic model of chemotactic aggregation}
\label{sec_stoc}

\subsection{Generalized Langevin equations}
\label{sec_lang}

We shall introduce a model of chemotactic aggregation generalizing the
Keller-Segel model (\ref{intro1})-(\ref{intro2}). For biological
systems, the number of constituents is not necessarily large so that
it may be relevant to return to a ``corpuscular'' description of the
system and introduce an equation of motion for each particle ($=$
cell). This type of ``microscopic'' approach has been previously
considered by Schweitzer \& Schimansky-Geier \cite{schweitzer}, Stevens
\cite{stevens} and Newman \& Grima \cite{grima} (see also related work by the authors \footnote{Chavanis {\it et al.}
\cite{crrs,hb,virial} studied stochastic models of Brownian particles
with long-range interactions where the motion of the individuals is
described by stochastic equations of the form (\ref{fluct18}) coupled
by a {\it binary} potential of interaction $u(|{\bf r}_{i}-{\bf
r}_{j}|)$ instead of the more complicated potential $c$, solution of
Eq.  (\ref{fluct2}), depending on the past history of the system
(memory terms are, however, considered in \cite{crrs,cban}). These
models describe, for example, self-gravitating Brownian particles
\cite{virial} and a simplified chemotactic model where
Eq. (\ref{fluct2}) is replaced by $\Delta
c-k_{0}^{2}c=-\lambda\sum_{\alpha=1}^{N}\delta({\bf r}-{\bf
r}_{\alpha}(t))$. This simplification is valid in a limit of large
diffusivity of the chemical (see Appendix \ref{sec_ld}) so that
$\partial c/\partial t$ can be neglected. In the mean field
approximation valid for $N\rightarrow +\infty$ in a proper
thermodynamic limit \cite{hb}, these models reduce to the
Smoluchowski-Poisson system \cite{crs} or to the Keller-Segel model
(\ref{intro1})- (\ref{intro2}) of chemotaxis where Eq. (\ref{intro2})
is replaced by $\Delta c-k_{0}^{2}c=-\lambda\rho$ \cite{degrad}. Note
that microscopic models yielding the regularized Keller-Segel model
(\ref{add1})-(\ref{add2}) have also been introduced in
\cite{degrad}.}). They describe the motion of individuals by 
a stochastic equation of the form
\begin{equation}
\label{fluct18} {d{\bf r}_{\alpha}\over dt}=\chi\nabla_{\alpha}c+\sqrt{2D_*}{\bf R}_{\alpha}(t).
\end{equation}
The first term in the r.h.s. is the chemotactic drift to which the
particles are submitted (the coefficient $\chi$ plays the role of a
mobility). The second term is a stochastic term where ${\bf
R}_{\alpha}(t)$ is a white noise satisfying $\langle {\bf
R}_{\alpha}(t)\rangle ={\bf 0}$ and $\langle
{R}_{i,\alpha}(t){R}_{j,\beta}(t')\rangle
=\delta_{ij}\delta_{\alpha,\beta}\delta(t-t')$ (where $\alpha=1,...,N$
refer to the particles and $i=1,...,d$ to the space coordinates) and
$D_{*}$ is a diffusion coefficient. The diffusion, that is observed
for several biological organisms, can have different origins depending
on the system under consideration \cite{perthame}. In the case of
small organisms moving in a fluid (matrigel), it can be due to the
repeated impact of the molecules of the fluid on the particles like in
ordinary Brownian motion for colloidal suspensions. In other cases, it
can be due to the properties of motion of the particles
themselves. For example, bacteria like {\it Escherichia coli} are
equipped with flagella and are self-propelled. When rotated
counterclockwise, the flagella act as a propellor and the bacterium
moves along straight line (``run''). Suddenly, the flagella rotate
clockwise and the bacterium stops to choose a new direction at random
(``tumble''). It continues in that direction for a while until the
next tumble. Therefore, the bacteria experience a random motion of
their own. At the simplest level of description, this motion can be
modelled by a stochastic term like in Eq. (\ref{fluct18}).  These
stochastic equations describe the motion of each of the $N$ particles
of the colony. As indicated above, $N$ is not necessarily large so it
may be of interest to treat the bacterial colony as a discrete system
of particles. By contrast, the chemical that is secreted is usually
described as a continuous field. Therefore, the evolution of the
concentration of the chemical is governed by an equation of the form
\begin{equation}
\label{fluct2}{\partial c\over\partial t}=-kc+D_{c}\Delta c+h\sum_{\alpha=1}^{N}\delta({\bf r}-{\bf r}_{\alpha}(t)).
\end{equation}
Equations (\ref{fluct18})-(\ref{fluct2}) have been studied in
\cite{schweitzer,stevens,grima,hb}. In the mean-field approximation,
they return the usual Keller-Segel model
(\ref{intro1})-(\ref{intro2}).

We shall generalize the model (\ref{fluct18})-(\ref{fluct2}) in two
respects. First of all, there exists biological systems for which the
inertia of the particles has to be taken into account
\cite{gamba}. This ``inertia'' means that they do not respond
immediately to the chemotactic drift. We propose therefore to describe
the motion of each individual of the biological population by a
stochastic equation of the form
\begin{equation}
\label{fluct1}{d{\bf r}_{\alpha}\over dt}={\bf v}_{\alpha}, \qquad {d{\bf v}_{\alpha}\over dt}=-\xi {\bf v}_{\alpha}+\nabla_{\alpha}c+\sqrt{2D}{\bf R}_{\alpha}(t).
\end{equation}
The first term in the r.h.s. is an ``effective'' friction force, the
second term is a force that models the chemotactic attraction due to
the chemical $c$ and the last term is a random force. The friction
force takes into account the fact that the velocity of the particles
has the tendency to be directed along the concentration gradient
$\nabla c$. This is exactly the case when $\xi\rightarrow +\infty$. In
this strong friction limit, we recover the overdamped model
(\ref{fluct18}) with $\chi=1/\xi$ and $D_{*}=D/\xi^{2}$. For finite
values of $\xi$, the velocity will take a (relaxation) time $\tau \sim
\xi^{-1}$ to get aligned with the concentration gradient. This is how
an inertial effect is introduced in the model. The term $-\xi {\bf v}$
can also represent a physical friction of the organisms against a
fixed matrigel. The stochastic equations (\ref{fluct1}), coupled
self-consistently to the field equation (\ref{fluct2}), describe the
motion of each of the $N$ particles of the colony. This completely
discrete model of chemotactic aggregation, which takes into account
statistical correlations between the particles, is developed in Appendix
\ref{sec_manybody}. An exact equation for the single-cell probability
distribution is derived and it is shown precisely how a mean field
approximation can be implemented in the theory.  In the mean-field
approximation, passing to a hydrodynamical description, we obtain the
model (\ref{intro3ht})-(\ref{intro5ht}) with a linear (isothermal)
equation of state $p=\rho T_{eff}=D\rho/\xi$.

In order to describe more general situations where the equation of
state $p=p(\rho)$ is nonlinear, we shall consider a generalized class
of stochastic equations where the diffusion coefficient explicitly
depends on the distribution function  $f({\bf r},{\bf v},t)$ of particles
in phase space.  These generalized stochastic equations have been introduced in
\cite{borland,gen,next,frank}. They are associated with nonlinear 
Fokker-Planck equations and lead to a notion of ``generalized
thermodynamics''. The possibility to apply this type of equations to
the chemotactic problem was proposed in \cite{gen}. In order to
simplify the formalism, we shall make a mean-field approximation since
the start and describe the motion of each individual of the biological
population by a stochastic equation of the form
\begin{equation}
\label{lang1} {d{\bf v}\over dt}=-\xi{\bf v}+\nabla S(c)+\sqrt{2D(f)}{\bf
R}(t),
\end{equation}
where the mean field force is determined by the equation
\begin{equation}
\label{lang2}{\partial c\over\partial t}=-k(c)c+h(c)\rho+D_{c}\Delta
c,
\end{equation}
where $\rho({\bf r},t)=\langle\sum_{i}\delta({\bf r}-{\bf
r}_{i}(t))\rangle$ is the {\it smooth} local density of cells (the
brackets denote an average over the noise). This amounts to replacing
the exact density $\rho_{ex}({\bf r},t)=\sum_{i}\delta({\bf r}-{\bf
r}_{i}(t))$ in Eq. (\ref{fluct2}) by the smooth density $\rho({\bf
r},t)=\langle\sum_{i}\delta({\bf r}-{\bf r}_{i}(t))\rangle$. This is
how the mean-field approximation is introduced in the model. For sake
of generality, we have allowed the coefficients $k$ and $h$ to depend
on the concentration $c$ and we have written the chemotactic force as
the gradient of a function $S(c)$ of the concentration.  Equations
(\ref{lang1})-(\ref{lang2}) will be our starting point for the
chemotactic problem. This model, or the discrete model (\ref{fluct1}),
is similar to the model of self-gravitating Brownian particles
introduced in \cite{crs,lang,virial}. The main difference, beyond the
context, is that the Poisson equation in gravity is replaced by the
more general field equation (\ref{lang2}).

\subsection{Nonlinear mean field Fokker-Planck equations}
\label{sec_fp}

We shall now derive kinetic and hydrodynamic equations associated
with the stochastic model (\ref{lang1})-(\ref{lang2}). Using
standard methods of Brownian theory, we find that the evolution of
the distribution function $f({\bf r},{\bf v},t)$ of the system is
described by a generalized mean field Fokker-Planck equation of the
form
\begin{equation}
\label{fp1} {\partial f\over\partial t}+{\bf v}\cdot {\partial
f\over\partial {\bf r}}+\nabla S(c)\cdot {\partial f\over\partial {\bf
v}} ={\partial\over\partial {\bf v}}\cdot \biggl\lbrack
{\partial \over\partial {\bf v}}(D(f)f)+\xi f{\bf
v}\biggr\rbrack.
\end{equation}
We have considered a relatively general class of stochastic processes
(\ref{lang1}) where the diffusion coefficient $D(f)$ can depend on the
distribution function.  This dependence can take into account
microscopic constraints of various origin (``hidden'' constraints)
that affect the dynamics at small scales. These generalized stochastic
processes are associated with a notion of generalized thermodynamics
\cite{frank} (see also Appendix \ref{sec_fe}). The case where $D(f)$ 
is a power law has been first
considered by Borland \cite{borland} in connection with Tsallis
generalized thermodynamics \cite{tsallis}. The case where $D(f)$ is
arbitrary has been considered by Chavanis \cite{gen} in connection
with nonlinear Fokker-Planck (NFP) equations associated with more general
forms of entropic functionals than the Tsallis entropy. If we write
the diffusion coefficient in the form $D(f)=Df\lbrack C(f)/f\rbrack'$,
where $C(f)$ is a convex function (i.e. $C''>0$), the nonlinear
Fokker-Planck equation (\ref{fp1}) can be rewritten \cite{gen}:
\begin{equation}
\label{fp2} {\partial f\over\partial t}+{\bf v}\cdot {\partial
f\over\partial {\bf r}}+\nabla S(c)\cdot {\partial f\over\partial {\bf
v}} ={\partial\over\partial {\bf v}}\cdot \biggl\lbrack
DfC''(f){\partial f\over\partial {\bf v}}+\xi f{\bf
v}\biggr\rbrack.
\end{equation}
This type of nonlinear Fokker-Planck equations can also be derived
from a master equation and a Kramers-Moyal expansion by allowing the
transition probabilities to depend on the occupation number in the
initial and arrival states as done in Kaniadakis
\cite{kaniadakis}. The case of a normal diffusion $D(f)=D$ corresponds
to $C(f)=f \ln f$ leading to the usual Kramers equation
(\ref{fluct15}) associated with the Boltzmann statistics. The choice
$D(f)=Df^{q-1}$ corresponds to $C(f)=(f^{q}-f)/(q-1)$ leading to the
polytropic Kramers equation
\cite{pp,gen,lang,logotropes} associated with the Tsallis statistics (these examples will be worked out explicitly in Sec. \ref{sec_gks}).

The stationary solution of Eq. (\ref{fp2}), which cancels both the
``collision'' term (r.h.s.) and the advection term (l.h.s.), is given by (see Appendix \ref{sec_fe}):
\begin{equation}
\label{fp3} C'(f_{eq})=-\beta\left \lbrack \frac{v^{2}}{2}-S(c)\right \rbrack-\alpha,
\end{equation}
where $\beta=1/T_{eff}$ is an effective inverse temperature satisfying
a generalized Einstein relation $\beta={\xi}/{D}$ and $\alpha$ is a
constant of integration.  Since $C$ is convex, the above relation can
be reversed. Then, we find that $f_{eq}({\bf r},{\bf
v})=F(\beta\epsilon+\alpha)=f_{eq}(\epsilon)$ where
$F(x)=(C')^{-1}(-x)$. Since $f_{eq}'(\epsilon)=-\beta/C''(f_{eq})$ and
$\beta>0$, we find that $f_{eq}(\epsilon)$ is a decreasing function of
the individual energy $\epsilon=v^{2}/2-S(c)$. The case of normal diffusion
$D(f)=D$ leads to the Maxwell-Boltzmann distribution $f_{eq}=A
e^{-\beta\epsilon}$. The case of anomalous diffusion $D(f)=Df^{q-1}$
leads to the Tsallis distribution $f_{eq}=\left\lbrack \lambda-\beta
(q-1)\epsilon/q\right\rbrack^{1/(q-1)}$ \cite{gen}.

\subsection{Damped hydrodynamic equations}
\label{sec_j}

We shall now derive the moments equations issued from the
generalized Fokker-Planck equation (\ref{fp2}). Defining the density
and the local velocity by
\begin{equation}
\label{j1} \rho=\int f\,d{\bf v}, \qquad \rho{\bf u}=\int f{\bf
v}\,d{\bf v},
\end{equation}
and integrating Eq.~(\ref{fp2}) on velocity, we get the continuity equation
\begin{equation}
\label{j2} {\partial\rho\over\partial t}+\nabla\cdot (\rho{\bf u})=0.
\end{equation}
Next, multiplying Eq.~(\ref{fp2}) by ${\bf v}$ and integrating on velocity, we obtain
\begin{equation}
\label{j3} {\partial\over\partial t}(\rho
u_{i})+{\partial\over\partial x_{j}}(\rho u_{i}u_{j})= -{\partial
P_{ij}\over\partial x_{j}}+\rho S'(c){\partial c\over\partial
x_{i}}-\xi\rho u_{i},
\end{equation}
where we have defined the ``pressure'' tensor
\begin{equation}
\label{j4}P_{ij}=\int fw_{i}w_{j}\,d{\bf v},
\end{equation}
where ${\bf w}={\bf v}-{\bf u}$ is the relative velocity.  Using the
continuity equation, Eq.~(\ref{j3}) can be rewritten
\begin{equation}
\label{j5} \rho\biggl ({\partial u_{i}\over\partial
t}+u_{j}{\partial u_{i}\over\partial x_{j}}\biggr )=-{\partial
P_{ij}\over\partial x_{j}}+\rho S'(c){\partial c\over\partial
x_{i}}-\xi\rho u_{i}.
\end{equation}

By taking the successive moments of the velocity, we can obtain a
hierarchy of hydrodynamic equations. Each equation of the
hierarchy involves the moment of next order.  If we are sufficiently
close to equilibrium, it makes sense to close the hierarchy of
equations by using a condition of local thermodynamic equilibrium
(L.T.E.). We shall thus evaluate the pressure tensor Eq.~(\ref{j4})
with the distribution function $f_{L.T.E}({\bf r},{\bf v},t)$
defined by the relation (see Appendix \ref{sec_km}):
\begin{equation}
\label{j6} C'(f_{L.T.E.})=-\beta\biggl \lbrack {w^{2}\over
2}+\lambda({\bf r},t)\biggr \rbrack.
\end{equation}
The function $\lambda({\bf r},t)$ is implicitly related to the
density by writing
\begin{equation}
\label{j7}\rho({\bf r},t)=\int f_{L.T.E.} \,d{\bf v}=\rho\lbrack
\lambda({\bf r},t)\rbrack.
\end{equation}
Using the condition
Eq.~(\ref{j6}) of local thermodynamic equilibrium, the
pressure tensor Eq.~(\ref{j4}) can be written
$P_{ij}=p\delta_{ij}$ with
\begin{equation}
\label{j8} p({\bf r},t)={1\over d}\int f_{L.T.E.} w^{2}\,d{\bf
w}=p\lbrack \lambda({\bf r},t)\rbrack.
\end{equation}
The pressure is a function $p=p(\rho)$ of the density which is
entirely specified by the function $C(f)$, by eliminating $\lambda$
from the relations Eq.~(\ref{j7}) and Eq.~(\ref{j8}). This defines a
barotropic gas. For example, the case of normal diffusion $D(f)=D$
leads to a Maxwellian distribution
$f_{L.T.E.}=(\beta/2\pi)^{d/2}\rho({\bf r},t){\rm exp}(-\beta
w^{2}/2)$ and a linear equation of state $p=\rho T_{eff}$ as for an
isothermal gas. The case of anomalous diffusion $D(f)=Df^{q-1}$ leads
to a Tsallis distribution $f_{L.T.E.}=\left\lbrack \mu({\bf
r},t)-\beta ((q-1)/q)w^{2}/2\right\rbrack^{1/(q-1)}$ and a power law
equation of state $p=K \rho^{\gamma}$ (with $\gamma=1+1/n$ and
$n=d/2+1/(q-1)$) as for a polytropic gas \cite{lang}. Substituting the
result $P_{ij}=p(\rho)\delta_{ij}$ in Eq. (\ref{j5}) and collecting
the other constitutive equations, we obtain a hydrodynamic model of
the form:
\begin{equation}
\label{j9} {\partial\rho\over\partial t}+\nabla\cdot (\rho{\bf
u})=0,
\end{equation}
\begin{equation}
\label{j10} {\partial {\bf u}\over\partial t}+({\bf u}\cdot
\nabla){\bf u}=-{1\over\rho}\nabla p+\nabla S(c)-\xi {\bf u},
\end{equation}
\begin{equation}
\label{j11}{\partial c\over\partial t}=-k(c)c+h(c)\rho+D_{c}\Delta c.
\end{equation}
The damped barotropic Euler equations (\ref{j9})-(\ref{j11}) are
interesting as they connect hyperbolic models to parabolic models
\cite{gen}. For $\xi=0$, we recover the hydrodynamic model (\ref{intro3})-(\ref{intro5}) introduced by Gamba {\it et al.} \cite{gamba} 
\footnote{In fact, the derivation of the damped barotropic equations (\ref{j9})-(\ref{j11}) from the kinetic theory developed in this section implicitly assumes that the friction force $-\xi {\bf u}$ is sufficiently large. Therefore, the limit $\xi\rightarrow 0$ is not really justified. Indeed, when $D,\xi\rightarrow 0$, Eq. (\ref{fp1}) reduces to the Vlasov equation and the L.T.E condition (\ref{j6}) is not realized or takes a long time to establish itself. In that collisionless regime, the system can undergo a process of ``violent relaxation'' driven by mean-field effects, like in astrophysics for the Vlasov-Poisson system \cite{csr}. In Appendix \ref{sec_hyp}, we present  an alternative  kinetic theory, based on other assumptions, leading to the hyperbolic model  (\ref{intro3})-(\ref{intro5})
proposed by Gamba {\it et al.} \cite{gamba}.}. Alternatively, for
$\xi\rightarrow +\infty$, we can formally neglect the inertial term in
Eq. (\ref{j10}) so that the velocity field is given by
\begin{equation}
\label{j12} {\bf u}=-{1\over\xi\rho}(\nabla
p-\rho S'(c)\nabla c)+O(\xi^{-2}).
\end{equation}
Substituting this drift term in the continuity equation (\ref{j9}), we obtain the drift-diffusion equation
\begin{equation}
{\partial\rho\over\partial t}=\nabla\cdot \biggl \lbrack
\frac{1}{\xi}\left (\nabla p-\rho S'(c)\nabla c\right
)\biggr\rbrack, \label{j13}
\end{equation}
which is a generalization of the Keller-Segel model. The usual
Keller-Segel model is recovered for a normal diffusion $D(f)=D$
leading to $p=\rho T_{eff}=D\rho/\xi$. We make the link with Eq.
(\ref{intro1}) by setting $D_*=D/\xi^2$, $\chi=1/\xi$ and $S(c)=c$.
The case of a polytropic equation of state $p=K\rho^{\gamma}$
associated to the Tsallis statistics has been studied in
\cite{lang,logotropes}.

It should be stressed that the damped Euler equations
(\ref{j9})-(\ref{j11}) remain heuristic because their derivation is
based on the Local Thermodynamic Equilibrium (L.T.E.) condition
(\ref{j6}) which is not rigorously justified. However, using a
Chapman-Enskog expansion, it is shown in \cite{lemou} that the
generalized Smoluchowski equation (\ref{j13}) is {\it exact} in the
limit $\xi\rightarrow +\infty$ (or, equivalently, for times $t\gg
\xi^{-1}$).  The generalized Smoluchowski equation can also be
obtained from the moments equations of the generalized Kramers
equation by closing the hierarchy in the limit $\xi\rightarrow
+\infty$ (see \cite{cban} and Appendix \ref{sec_sf}). 

\subsection{A generalized Keller-Segel model}
\label{sec_gks}

In this section, we discuss an explicit example to illustrate our
general formalism. We consider a stochastic process of the form
\begin{equation}
\label{gks1} {d{\bf v}\over dt}=-\xi{\bf v}+\nabla S(c)+\sqrt{2D}f^{(q-1)/2}{\bf
R}(t),
\end{equation}
where the mean field force is determined by
Eq. (\ref{lang2}). Comparing with Eq. (\ref{lang1}), we find that the
diffusion coefficient is given by $D(f)=Df^{q-1}$. As indicated
previously, this leads to a situation of anomalous diffusion related
to the Tsallis statistics \cite{borland}.  For $q=1$, we recover the
standard Brownian model with a constant diffusion coefficient,
corresponding to a pure random walk (see Appendix
\ref{sec_manybody}). In that case, the sizes of the random kicks are
uniform and do not depend on where the particle happens to be. For
$q\neq 1$, the size of the random kicks changes, depending on the
phase-space distribution of the particles around the ``test''
particle. A particle which is in a state $({\bf r},{\bf v})$ that is
highly populated [large $f({\bf r},{\bf v},t)$] will tend to have
larger kicks if $q>1$ and smaller kicks if $q<1$. Since the
microscopics depends on the actual density in phase space, this
creates a bias in the ergodic behavior of the system.  The nonlinear
Fokker-Planck equation associated with the stochastic process
(\ref{gks1}) is the polytropic Kramers equation
\begin{equation}
\label{gsk2} {\partial f\over\partial t}+{\bf v}\cdot {\partial
f\over\partial {\bf r}}+\nabla S(c)\cdot {\partial f\over\partial {\bf
v}} ={\partial\over\partial {\bf v}}\cdot \biggl ( D {\partial
f^{q}\over\partial {\bf v}}+\xi f{\bf v}\biggr ).
\end{equation} 
For $q=1$, we recover the classical Kramers equation (\ref{fluct15})
which can be deduced from the $N$-body stochastic process
(\ref{fluct1}) by using a mean field approximation (see Appendix
\ref{sec_manybody}). For $q\neq 1$, the NFP equation 
(\ref{gsk2}) can be deduced from the mean field stochastic process
(\ref{gks1}). The associated Lyapunov functional (\ref{fe2}) is
explicitly given by
\begin{eqnarray}
\label{gsk3}
F[f]=\int f\frac{v^{2}}{2}d{\bf r}d{\bf v}+\frac{1}{2h}\int \left\lbrack D_{c}(\nabla c)^{2}+kc^{2}\right\rbrack d{\bf r}
-\int \rho c d{\bf r}+\frac{T_{eff}}{q-1}\int (f^{q}-f)d{\bf r}d{\bf v},
\end{eqnarray}
where $T_{eff}=1/\beta=D/\xi$. It can be viewed as an effective
generalized free energy $F=E-T_{eff}S$ associated with the Tsallis
entropy $S_{q}=-(1/(q-1))\int (f^{q}-f)d{\bf r}d{\bf v}$. It satisfies
$\dot F\le 0$ which is the version of the $H$-theorem in the canonical
ensemble where the temperature $T_{eff}$ is fixed instead of the
energy. The steady state of the NFP equation (\ref{gsk2}) is the
polytropic (Tsallis) distribution
\begin{equation}
f_{eq}=\left\lbrack \mu-\frac{\beta(q-1)}{q}\epsilon\right\rbrack^{1/(q-1)}, \label{gsk4}
\end{equation}
where $\epsilon=v^{2}/2-S(c)$ is the energy per particle. This
distribution extremizes the free energy (\ref{gsk3}) at fixed
mass. Furthermore, it is linearly dynamically stable with respect to
the NFP equation (\ref{gsk2}) if, and only if, it is a minimum of $F$
at fixed $M$ \cite{gen}. The index $n$ of the polytrope (defined by
Eq. (\ref{gsk10}) below) is related to the parameter $q$ by the
relation
\begin{equation}
\label{gsk5} n=\frac{d}{2}+\frac{1}{q-1}.
\end{equation}
The isothermal (Boltzmann) distribution function
$f_{eq}=Ae^{-\beta\epsilon}$, corresponding to normal diffusion, is
recovered in the limit $q\rightarrow 1$, i.e. $n\rightarrow
+\infty$. In the following, we shall consider $q>0$ so that $C$ is
convex. Since $\beta>0$, this implies that $f(\epsilon)$ is decreasing
(see Sec. \ref{sec_fp}). There are two cases to consider. For $q>1$,
i.e. $n>d/2$, the distribution (\ref{gsk4}) can be written
\begin{equation}
\label{gsk6} f=A(\epsilon_{m}-\epsilon)^{1/(q-1)}\quad ({\rm case}\ 1)
\end{equation}
where $A=\lbrack \beta(q-1)/q\rbrack^{1/(q-1)}$ and
$\epsilon_{m}=q\mu/\lbrack \beta (q-1)\rbrack$. It has a compact
support since $f$ is defined only for $\epsilon\le \epsilon_{m}$. For
$\epsilon\ge \epsilon_{m}$, we set $f=0$ (for $q=+\infty$,
i.e. $n=d/2$, $f$ is the Heaviside function).  For $q<1$ the
distribution can be written
\begin{equation}
\label{gsk7} f=A(\epsilon_{0}+\epsilon)^{-1/(1-q)}\quad ({\rm case}\ 2)
\end{equation}
where $A=\lbrack \beta(1-q)/q\rbrack^{-1/(1-q)}$ and
$\epsilon_{0}=q\mu/\lbrack \beta (1-q)\rbrack$. It is defined for
all energies. For large velocities, it behaves like $f\sim
v^{2n-d}$. Therefore, the density and the pressure are finite only
for $n<-1$, i.e. $d/(d+2)<q<1$. Therefore the range of allowed
parameters are
\begin{equation}
\label{gsk8} q>1, \qquad n>\frac{d}{2}\quad ({\rm case}\ 1)
\end{equation}
\begin{equation}
\label{gsk9} \frac{d}{d+2}<q<1, \qquad n<-1 \quad ({\rm case}\ 2)
\end{equation}
From Eq. (\ref{gsk4}), or the alternative forms (\ref{gsk6}) and
(\ref{gsk7}), we can compute the density $\rho=\int f d{\bf
v}=\rho[c({\bf r})]$ and the pressure $p=(1/d)\int f v^{2}d{\bf
v}=p[c({\bf r})]$ at equilibrium (see Sec. 3.6 of
\cite{interpretation} for details). Then, eliminating the
concentration $c({\bf r})$ between these expressions, we obtain the
polytropic equation of state
\begin{equation}
\label{gsk10} p=K\rho^{\gamma}, \qquad \gamma=1+{1\over n}.
\end{equation}
For $n>d/2$ (case 1), the polytropic constant is
\begin{equation}
\label{gsk11} K=\frac{1}{n+1}\left\lbrack A S_{d}
2^{\frac{d}{2}-1}\frac{\Gamma\left (d/2\right )\Gamma\left
(1-d/2+n\right )}{\Gamma(1+n)}\right \rbrack^{-1/n},
\end{equation}
and for $n<-1$ (case 2), we have
\begin{equation}
\label{gsk12} K=-\frac{1}{n+1}\left\lbrack A S_{d}
2^{\frac{d}{2}-1}\frac{\Gamma\left (d/2\right )\Gamma\left (-n\right
)}{\Gamma(d/2-n)}\right \rbrack^{-1/n},
\end{equation}
where $\Gamma(x)$ is the Gamma function. For $q=1$, the polytropic
equation of state (\ref{gsk10}) reduces to  the ``isothermal'' (linear)
equation of state $p=\rho T_{eff}$. These equations of state have been
obtained from the distribution (\ref{gsk4}) valid at
equilibrium. However, the same equations of state are obtained from the
L.T.E condition (\ref{j6}), using Eqs. (\ref{j7}) and (\ref{j8}) and
$C(f)=(f^{q}-f)/(q-1)$, or from the distribution function (\ref{sf2})
valid in the strong friction limit $\xi\rightarrow +\infty$, using
Eqs. (\ref{sf3}) and (\ref{sf4}). Then, the damped hydrodynamic
equations corresponding to the model (\ref{gks1}) are
Eqs. (\ref{j9})-(\ref{j11}) with the equation of state
(\ref{gsk10}). On the other hand, in the strong friction limit, the
drift-diffusion equation (\ref{j13}) reduces to the polytropic
Smoluchowski equation
\begin{equation}
{\partial\rho\over\partial t}=\nabla\cdot \biggl \lbrack
\frac{1}{\xi}\left (K\nabla \rho^{\gamma}-\rho S'(c)\nabla c\right
)\biggr\rbrack. \label{gsk13}
\end{equation}
The associated Lyapunov functional (\ref{fe28}) is explicitly given by
\begin{eqnarray}
\label{gsk14}
F[\rho]={K\over\gamma -1}\int (\rho^{\gamma}-\rho) \ d{\bf
r}+\frac{1}{2h}\int \left\lbrack D_{c}(\nabla c)^{2}+kc^{2}\right\rbrack d{\bf r}
-\int \rho c \, d{\bf r}.
\end{eqnarray}
It can be viewed as an effective generalized free energy $F=E-KS$
associated with the Tsallis entropy $S_{\gamma}=-(1/(\gamma-1))\int
(\rho^{\gamma}-\rho)d{\bf r}$ where $K$ plays the role of an effective
temperature and $\gamma$ the role of the Tsallis $q$-parameter. It
satisfies $\dot F\le 0$ which is the version of the $H$-theorem in the
canonical ensemble where the ``polytropic temperature'' $K$ is fixed instead of
the energy. The steady state of the generalized  Smoluchowski equation (\ref{gsk13}) is the polytropic (Tsallis) distribution in physical space 
\begin{equation}
\label{gsk15} \rho_{eq}=\biggl\lbrack \lambda+{\gamma-1\over
K\gamma}S(c)\biggr\rbrack^{1\over\gamma-1}.
\end{equation}
This distribution extremizes the free energy (\ref{gsk14}) at fixed
mass. Furthermore, it is linearly dynamically stable with respect to
the generalized Smoluchowski equation (\ref{gsk13}) if and only if it
is a minimum of $F$ at fixed $M$.  Of course,
Eq. (\ref{gsk15}) can be deduced from Eq. (\ref{gsk4}). Equation
(\ref{gsk13}) can be interpreted as a NFP equation of the form
considered in \cite{gen} with
$C(\rho)=(\rho^{\gamma}-\rho)/(\gamma-1)$. It can be obtained directly
from the generalized stochastic process (see Appendix
\ref{sec_prim})
\begin{equation}
\label{gsk16} 
\frac{d{\bf r}}{dt}=\frac{1}{\xi}\rho\nabla S(c)+\sqrt{\frac{2K}{\xi}}\rho^{(\gamma-1)/2}{\bf R}(t).
\end{equation}
This stochastic process generalizes Eq. (\ref{fluct18}) and provides
another justification of the polytropic Smoluchowski equation
(\ref{gsk13}). The interpretation of the noise in Eq. (\ref{gsk16}),
depending on the density $\rho({\bf r},t)$, is similar to that given
after Eq. (\ref{gks1}) although the process takes place in physical
space instead of phase space.

Finally, we note the remarkable feature that a polytropic distribution
$f_{eq}(\epsilon)$ [see Eq. (\ref{gsk4})] with index $q$ in phase
space yields a polytropic distribution $\rho_{eq}(c)$ [see
Eq. (\ref{gsk15})] with index $\gamma$ in physical space (using
Eqs. (\ref{gsk5}) and (\ref{gsk10}), the indices are related to each
other by $\gamma=1+2(q-1)/\lbrack 2+d(q-1)\rbrack$). In this sense,
polytropic (Tsallis) distributions are ``stable'' laws. Apparently,
these are the only ones enjoying the property that $f_{eq}(\epsilon)$
and $\rho_{eq}(c)$ have the same form. By comparing Eqs. (\ref{gsk4})
and (\ref{gsk15}) or Eqs. (\ref{gsk3}) and (\ref{gsk14}), we note that
$K$ plays the same role in physical space as the temperature
$T_{eff}=1/\beta$ in phase space. It is sometimes called a
``polytropic temperature''. We also note that for $q>1$, we have
$\gamma>1$ so that the model is hyper-diffusive in phase space and
physical space. For $q<1$, we have $\gamma<1$ so that the model is
sub-diffusive in phase space and physical space. For $q=1$, we have
$\gamma=1$ so that Eq. (\ref{gsk13}) reduces to the ordinary
Smoluchowski equation. This yields the standard Keller-Segel model
\cite{ks}. For $q\neq 1$, or $\gamma\neq 1$, we obtain a
generalized Keller-Segel model that can take into account non-ideal
effects giving rise to anomalous diffusion. This model has been
studied in \cite{lang,logotropes,masscrit}.

\subsection{Numerical simulations}
\label{sec_ns}

In this section, we restrict ourselves to normal diffusion and we
present numerical simulations of the $N$-body system
(\ref{fluct1})-(\ref{fluct2}) in a simplified setting. In the equation
(\ref{fluct2}) for the evolution of the secreted chemical, we set
$h=\lambda D_{c}$ and consider a limit of large diffusivity
$D_c\rightarrow +\infty$ with $\lambda\sim 1$. In that limit, the
temporal derivative $\partial c/\partial t$ and the degradation term
$-kc$ can be neglected (see Appendix
\ref{sec_ld}). The model (\ref{fluct1})-(\ref{fluct2}) becomes
\begin{equation}
\label{ns1} {d{\bf v}_{\alpha}\over dt}=-\xi{\bf v}_{\alpha}+\nabla_{\alpha} c+\sqrt{2D}{\bf
R}_{\alpha}(t),
\end{equation}
\begin{equation}
\label{ns2}\Delta c=-\lambda(\rho_{ex}-\overline{\rho}),
\end{equation}
where $\rho_{ex}({\bf r},t)=\sum_{\alpha=1}^{N}\delta({\bf r}-{\bf r}_{\alpha}(t))$ is the exact density and $\overline{\rho}=N/V$ is the average density over the entire domain. When $D=\xi=0$, we obtain
\begin{equation}
\label{ns3} {d{\bf v}_{\alpha}\over dt}=\nabla_{\alpha} c,
\end{equation}
\begin{equation}
\label{ns4}\Delta c=-\lambda(\rho_{ex}-\overline{\rho}).
\end{equation}
These equations are similar  to the Newton equations for a
self-gravitating system where $-c$ plays the role of the gravitational
potential $\Phi$ and $\lambda$ the role of the gravitational constant
$S_{d} G$ ($S_{d}$ is the surface of a unit sphere in
$d$-dimensions). In cosmology, when we take into account the expansion
of the universe and work in a comoving frame, the usual Poisson
equation $\Delta\Phi=4\pi G\rho$ is replaced by an equation of the
form $\Delta\phi=4\pi G a(t)^{2}\lbrack \rho({\bf
x},t)-\rho_{b}(t)\rbrack$ where the density $\rho({\bf x},t)$ is
replaced by the deviation $\rho({\bf x},t)-{\rho}_{b}(t)$ to the mean
density \cite{peebles}. Furthermore, the equations of motion read
$(d/dt)(ma^{2}\dot {\bf x})=-m\nabla\phi$. If we consider timescales
over which the variation of the scale factor $a(t)$ can be neglected,
the equations of motion become isomorphic to
Eqs. (\ref{ns3})-(\ref{ns4}). Therefore, there exists interesting
analogies between the process of chemotaxis in biology and the
dynamics of self-gravitating systems.

\begin{figure} 
\vskip1cm
\centerline{
\psfig{figure=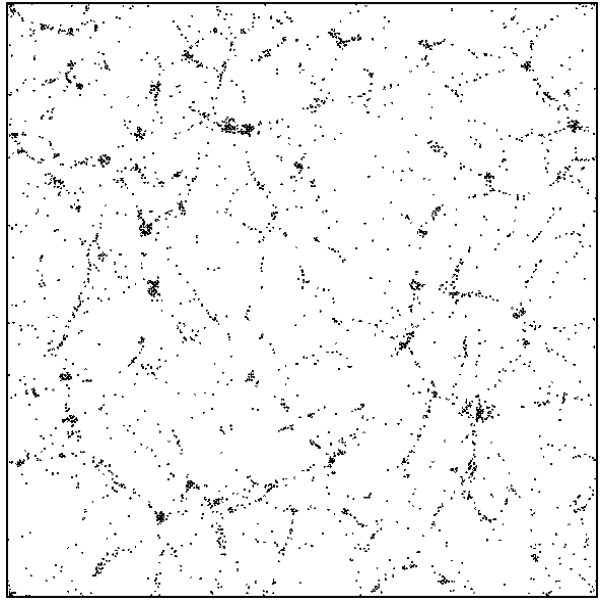,angle=0,height=8cm}} 
\caption{$N$-body simulation of the inertial model (\ref{ns3})-(\ref{ns4}) for $N=32768$ particles 
in a periodic box. The dynamical equations are isomorphic to the
Newton equations for a self-gravitating system in cosmology. They show
the formation of a network pattern with a filamentary structure. In
biology, this corresponds to the beginning of a vasculature. } 
\label{newton}
\end{figure}

\begin{figure}
\vskip1cm
\centerline{
\psfig{figure=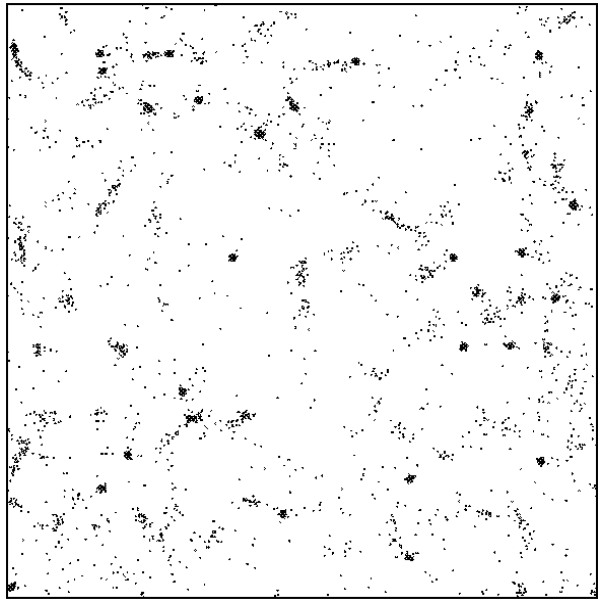,angle=0,height=8cm}}
\vskip0.2cm
\caption{$N$-body simulation of the overdamped model (\ref{ns5})-(\ref{ns6}) for $N=32768$ particles 
in a periodic box. The dynamical equations are isomorphic to the
Langevin equations for a self-gravitating Brownian gas in an
overdamped limit. They lead to point-wise blow-up forming ultimately
Dirac peaks.  In biology, this corresponds to a chemotactic collapse.} 
\label{brown}
\end{figure}

However, in biology, we are rather in a limit where inertial effects
are negligible or weak so that the (effective) friction coefficient $\xi$ is
relatively large. Therefore, the process of chemotaxis in biology is
more directly analogous to the dynamics of self-gravitating Brownian
particles studied in \cite{crs}. If we consider the strong friction
limit $\xi\rightarrow +\infty$, we obtain
\begin{equation}
\label{ns5} {\bf v}_{\alpha}=\frac{1}{\xi}\nabla_{\alpha} c+\sqrt{2D_*}{\bf R}_{\alpha}(t),
\end{equation}
\begin{equation}
\label{ns6}\Delta c=-\lambda (\rho_{ex}-\overline{\rho}).
\end{equation}
These equations are similar to those describing a
self-gravitating Brownian gas in an overdamped limit \cite{virial}.

The inertial model (\ref{ns3})-(\ref{ns4}) leads to fluid models of
a form related to Eqs. (\ref{intro3})-(\ref{intro5}) that are
hyperbolic. Alternatively, the overdamped model
(\ref{ns5})-(\ref{ns6}) leads to drift-diffusion equations of the form
(\ref{intro1})-(\ref{intro2}) that are parabolic.  As discussed in the
Introduction, parabolic models are known to lead to pointwise blow-up
while hyperbolic models generate network patterns
\cite{gamba,filbet}. We have performed direct numerical simulations
of the $N$-body systems (\ref{ns3})-(\ref{ns4}) and
(\ref{ns5})-(\ref{ns6}) in a periodic domain starting from a
statistically homogeneous distribution of particles. In order to
simulate a large number of particles, we have considered the dimension
$d=2$ which is relevant for biological populations.  The results of
the simulations are reported in Figs. \ref{newton} and
\ref{brown}. The uniform distribution of particles is unstable (this
is similar to the Jeans instability in astrophysics) and the particles
start to collapse and form aggregates. Note that the concentration can
be quite large although this is not always obvious on the figures
since the particles have fallen on each other. Density contrasts are
easier to see in continuous (fluid) models that automatically involve
a coarse-graining
\cite{gamba,filbet}. However, our $N$-body simulations also
show the formation of lines and filaments in the case of models with
inertia (Fig. \ref{newton}) and the absence (or reduction) of such
lines and the generation of pointwise blow-up in the case of
overdamped systems (Fig. \ref{brown}).

\section{Conclusion}

In this paper, we have introduced general kinetic and hydrodynamic
models of chemotactic aggregation. These models can be relevant to
describe the organization of social insects (swarms) like ants or the
morphogenesis of biological populations like bacteria, amoebae,
endothelial cells etc.  Starting from a microscopic model defined in
terms of $N$ coupled stochastic equations
(\ref{fluct1})-(\ref{fluct2}) or (\ref{lang1})-(\ref{lang2}), we have
derived a generalized mean field Fokker-Planck equation (\ref{fp2})
governing the evolution of the distribution function of the system in
phase space. By taking the successive moments of this kinetic equation
and closing the hierarchy by a local thermodynamic equilibrium
condition, we have derived a set of hydrodynamic equations
(\ref{j9})-(\ref{j11}) involving a damping term.  An interest of this
approach is to connect and generalize different models previously
introduced in the literature. In particular, the Keller-Segel model
(\ref{intro1})-(\ref{intro2}) and the generalized Keller-Segel model
(\ref{add1})-(\ref{add2}) are obtained in a limit of strong friction
$\xi\rightarrow +\infty$ and the hydrodynamic model introduced by
Gamba {\it et al.} (\ref{intro3})-(\ref{intro5}) corresponds formally
to a limit of low friction $\xi\rightarrow 0$ (in fact, the
justification of this model must be given separately as in Appendix
\ref{sec_hyp}). We have illustrated numerically the difference between
models with inertia (hyperbolic) leading to network patterns and
models without inertia (parabolic) leading to pointwise blow-up. We
have discussed the analogy between the chemotactic collapse in biology
and the gravitational collapse (Jeans instability) in astrophysics
(see also \cite{epjbjeans}).

In our kinetic model of Sec. \ref{sec_fp}, the dynamical evolution of
bacterial populations is described by generalized stochastic processes
(\ref{lang1}) and nonlinear mean field Fokker-Planck equations
(\ref{fp2}) similar to those arising in the context of generalized
thermodynamics \cite{gen,next,frank}. As a result, the hydrodynamic
equation (\ref{j10}) involves a {\it nonlinear} pressure term
$p(\rho)$. If we consider stochastic processes with normal diffusion
(\ref{fluct1}), we obtain ordinary Fokker-Planck equations
(\ref{fluct15}) leading to a linear equation of state $p(\rho)=\rho
T_{eff}$ (isothermal) with $T_{eff}=D/\xi$. In the strong friction
limit, this yields the Keller-Segel model (\ref{intro1}) which is
known to form singularities (Dirac peaks). In order to obtain the
regularized model (\ref{add1}), we need to modify the stochastic
equations.  This has been done here by letting the diffusion
coefficient depend on the local distribution of particles in phase
space (see Eq. (\ref{lang1})). This is a phenomelogical attempt to
take into account complicated microscopic constraints (like excluded
volume constraints, finite size effects, short-range interactions,...) 
that affect the dynamics of the particles at small scales. At the
level of the hydrodynamic equations, these constraints are modeled by
an effective pressure term $p=p(\rho)$ which replaces the usual term
$p=\rho T_{eff}$. An example of regularized Keller-Segel model where
the Dirac peaks are replaced by smooth density profiles (aggregates)
has been studied in
\cite{degrad}. In that case, the effective equation of state is
$p(\rho)=-\sigma_{0}T_{eff}\ln(1-\rho/\sigma_{0})$. For dilute systems
where the motion of an individual cell is not impeded by the other
cells, we have $\rho\ll \sigma_{0}$ and we recover the ``isothermal''
equation of state $p=\rho T_{eff}$. However, modifications arise when
the cells are compressed. Indeed, the equation of state departs from
the isothermal one when the density approaches the maximum allowable
density $\sigma_{0}$. In that case, instead of Dirac peaks, we form
flat cores with density $\rho \sim \sigma_{0}$. The nonlinear pressure
term $p(\rho)$ can also model a process of anomalous diffusion. This
is the case in particular for the stochastic model (\ref{gks1})
leading to a polytropic equation of state $p=K\rho^{\gamma}$. This
model can take into account effects of non-ergodicity and
nonextensivity. This leads to the Tsallis statistics that arises
when the dynamics has a fractal or multi-fractal phase space structure
\cite{borland}.  The corresponding generalized Keller-Segel
model (\ref{gsk13}) has been studied in detail in
\cite{lang,logotropes,masscrit}.

The next step is a detailed study of the models presented in this
paper. This study is of interest not only in mathematical biology but
also, at a more general level, in statistical mechanics. What we are
studying essentially is the {\it Dynamics and Thermodynamics of
Brownian Particles with Long-Range Interactions} \cite{hb}. We have
furthermore introduced a general class of stochastic equations with a
random force depending on the distribution function, forcing the
nonlinearity in the Fokker-Planck equation. This is associated with a
notion of generalized thermodynamics. Therefore, our model involves
both long-range forces \cite{houches} and generalized thermodynamics
(related to nonlinear Fokker-Planck equations) \cite{frank}, which are two
domains actively studied at the moment in statistical mechanics. Our
approach shows that these topics can have applications in biology. In the
overdamped limit, the dynamical equations have the form of
drift-diffusion equations corresponding to the Keller-Segel model or
generalizations of this model. These equations have been extensively
studied in the mathematical (see the review of Horstmann
\cite{horstmann}) and physical (see Chavanis \& Sire \cite{virial} and
references therein) literature. The more complicated study of kinetic
equations (\ref{fp2})-(\ref{lang2}) or hydrodynamic equations
(\ref{j9})-(\ref{j11}) should be considered in future works. The
connection with generalized forms of Cahn-Hilliard equations when the
potential of interaction is short-ranged (as noticed in
\cite{next,lemou}) should also be developed (see Appendix
\ref{sec_ch}).

\appendix

\section{Many-body theory of chemotactic aggregation}
\label{sec_manybody}

In this Appendix, we first develop an {\it exact} many-body theory of
Brownian particles in interaction. Then, we show how a mean field
approximation can be implemented in the problem. For simplicity, we
assume that each particle has a normal Brownian motion, i.e. the
diffusion coefficient $D$ is constant. Our approach follows the steps of
Newman \& Grima \cite{grima}. However, we take into account the inertia of the
particles while Newman \& Grima    consider an overdamped limit.

Basically, the dynamical evolution of $N$ Brownian particles in
interaction is described by stochastic Langevin equations of the form
(\ref{fluct1}). In the biological context, the field $c$ represents
the concentration of the chemical produced by the 
cells and it satisfies an equation of the form (\ref{fluct2}). We
define the single-cell probability distribution in phase space by
\begin{equation}
\label{fluct3}P_{\alpha}({\bf r},{\bf v},t)=\langle \delta({\bf r}-{\bf r}_{\alpha}(t))\delta({\bf v}-{\bf v}_{\alpha}(t))\rangle,
\end{equation}
where the brackets denote an average over the noise. Similarly the  two-cell probability distribution in phase space is
\begin{eqnarray}
\label{fluct4}P_{\alpha,\beta}({\bf r},{\bf v},t;{\bf r}',{\bf v}',t')=\langle \delta({\bf r}-{\bf r}_{\alpha}(t))\delta({\bf v}-{\bf v}_{\alpha}(t))
\delta({\bf r}'-{\bf r}_{\beta}(t'))\delta({\bf v}'-{\bf v}_{\beta}(t'))\rangle.
\end{eqnarray} 
Integrating Eq. (\ref{fluct2}), the concentration field can be
expressed in terms of the cell paths as
\begin{eqnarray}
\label{fluct5}c({\bf r},t)=h\int d{\bf r}'\int_{0}^{t}dt' G({\bf r}-{\bf r}',t-t')\sum_{\alpha}\delta({\bf r}'-{\bf r}_{\alpha}(t')),
\end{eqnarray}  
where the Green function for the chemical diffusion equation is
\begin{eqnarray}
\label{fluct6}G({\bf r},t)=(4\pi D_{c}t)^{-d/2}{\rm exp}\left\lbrack -{r^{2}\over 4D_{c}t}-kt\right\rbrack.
\end{eqnarray} 
Taking the time derivative of $P_{\alpha}$, we get
\begin{eqnarray}
\label{fluct7}{\partial P_{\alpha}\over\partial t}=-{\partial\over\partial {\bf r}}\cdot \langle {\dot{\bf r}}_{\alpha}(t) \delta({\bf r}-{\bf r}_{\alpha}(t))\delta({\bf v}-{\bf v}_{\alpha}(t))\rangle
 -{\partial\over\partial {\bf v}}\cdot \langle {\dot{\bf v}}_{\alpha}(t) \delta({\bf r}-{\bf r}_{\alpha}(t))\delta({\bf v}-{\bf v}_{\alpha}(t))\rangle . 
\end{eqnarray} 
Inserting the equations of motion (\ref{fluct1}) in
Eq. (\ref{fluct7}), we obtain
\begin{eqnarray}
\label{fluct8}{\partial P_{\alpha}\over\partial t}=-{\partial\over\partial {\bf r}}\cdot \langle {{\bf v}}_{\alpha}(t)\ \delta({\bf r}-{\bf r}_{\alpha}(t))\delta({\bf v}-{\bf v}_{\alpha}(t))\rangle
 +{\partial\over\partial {\bf v}}\cdot\langle \xi {{\bf v}}_{\alpha}(t)\ \delta({\bf r}-{\bf r}_{\alpha}(t))\delta({\bf v}-{\bf v}_{\alpha}(t))\rangle \nonumber\\ 
-{\partial\over\partial {\bf v}}\cdot \langle \nabla_{\alpha}c \ \delta({\bf r}-{\bf r}_{\alpha}(t))\delta({\bf v}-{\bf v}_{\alpha}(t))\rangle
-{\partial\over\partial {\bf v}}\cdot\langle \sqrt{2D} {\bf R}_{\alpha}(t) \ \delta({\bf r}-{\bf r}_{\alpha}(t))\delta({\bf v}-{\bf v}_{\alpha}(t))\rangle.  \end{eqnarray} 
The first two terms are straightforward to evaluate. The fourth term is the standard term that appears in deriving the Fokker-Planck equation for a pure random walk; it leads to a term proportional to the Laplacian of $P_{\alpha}$ in velocity space.  The third term can be evaluated by inserting the formal solution (\ref{fluct5}) in Eq. (\ref{fluct8}). This leads to the exact equation
\begin{eqnarray}
\label{fluct9}{\partial P_{\alpha}\over\partial t}+{\bf v}\cdot {\partial P_{\alpha}\over\partial {\bf r}}+h{\partial\over\partial {\bf v}}\cdot \int d{\bf r}'\int_{0}^{t}dt' \nabla G ({\bf r}-{\bf r}',t-t')\sum_{\beta}P_{\alpha,\beta}({\bf r},{\bf v},t;{\bf r}',t')\nonumber\\
={\partial\over\partial {\bf v}}\cdot \left \lbrack D{\partial P_{\alpha}\over\partial {\bf v}}+\xi P_{\alpha}{\bf v}\right\rbrack,  \qquad\qquad  
\end{eqnarray} 
where the statistical correlations are encapsulated in the two-cell distribution
\begin{eqnarray}
\label{fluct10}P_{\alpha,\beta}({\bf r},{\bf v},t;{\bf r}',t')=\langle \delta({\bf r}-{\bf r}_{\alpha}(t))\delta({\bf v}-{\bf v}_{\alpha}(t))
\delta({\bf r}'-{\bf r}_{\beta}(t'))\rangle.
\end{eqnarray} 
The mean field approximation is implemented by assuming that the two-cell distribution factorizes
\begin{eqnarray}
\label{fluct11}P_{\alpha,\beta}({\bf r},{\bf v},t;{\bf r}',t')=P_{\alpha}({\bf r},{\bf v},t)P_{\beta}({\bf r}',t').
\end{eqnarray} 
In that case, the preceding equation can be rewritten
\begin{eqnarray}
\label{fluct12}{\partial P_{\alpha}\over\partial t}+{\bf v}\cdot {\partial P_{\alpha}\over\partial {\bf r}}+h{\partial\over\partial {\bf v}}\cdot P_{\alpha}({\bf r},{\bf v},t)  \nabla \int d{\bf r}'\int_{0}^{t}dt' G ({\bf r}-{\bf r}',t-t')\sum_{\beta}P_{\beta}({\bf r}',t')\nonumber\\
={\partial\over\partial {\bf v}}\cdot \left \lbrack D{\partial P_{\alpha}\over\partial {\bf v}}+\xi P_{\alpha}{\bf v}\right\rbrack.  \qquad\qquad  
\end{eqnarray} 
We define the distribution function and the density by
\begin{eqnarray}
\label{fluct13}
f({\bf r},{\bf v},t)=\sum_{\alpha=1}^{N}P_{\alpha}({\bf r},{\bf v},t),
\end{eqnarray} 
\begin{eqnarray}
\label{fluct14}
\rho({\bf r},t)=\sum_{\alpha=1}^{N}P_{\alpha}({\bf r},t)=\int f({\bf r},{\bf v},t) d{\bf v}.
\end{eqnarray} 
Summing Eq. (\ref{fluct12}) over the cell index, we obtain a kinetic equation of the form
\begin{eqnarray}
\label{fluct15}{\partial f\over\partial t}+{\bf v}\cdot {\partial f\over\partial {\bf r}}+\nabla c \cdot {\partial f\over\partial {\bf v}}
={\partial\over\partial {\bf v}}\cdot \left \lbrack D{\partial f\over\partial {\bf v}}+\xi f{\bf v}\right\rbrack,  \qquad\qquad  
\end{eqnarray} 
where 
\begin{eqnarray}
\label{fluct16}
c({\bf r},t)=h\int d{\bf r}'\int_{0}^{t}dt' G ({\bf r}-{\bf r}',t-t')\rho({\bf r}',t'). 
\end{eqnarray}  
This smooth field satisfies the partial differential equation
\begin{equation}
\label{fluct17}{\partial c\over\partial t}=-kc+D_{c}\Delta c+h\rho.
\end{equation}
Note that Eq. (\ref{fluct15}) can be viewed as a mean field Kramers
equation \cite{hb}. For $D=\xi=0$, it becomes equivalent to the Vlasov
equation.

If we consider a limit of strong frictions (or large times $t\gg
\xi^{-1}$) in the stochastic equations (\ref{fluct1}), we obtain the overdamped model (\ref{fluct18}). These equations, coupled to
Eq.  (\ref{fluct2}) for the chemical concentration field have been
considered in \cite{grima}. The above procedure leads, in the
mean field approximation, to a kinetic equation of the form (\ref{intro1})
coupled to Eq. (\ref{fluct17}). This returns the ordinary
Keller-Segel model. Equation (\ref{intro1}) can be viewed as a
mean field Smoluchowski equation \cite{hb}.  It can also be obtained from the
mean field Kramers equation Eq. (\ref{fluct15}) by considering the
strong friction limit and using a Chapman-Enskog expansion
\cite{lemou}.  Therefore, our model (\ref{fluct1}) can be considered
as an extension of the overdamped model of \cite{grima} when inertial
effects are taken into account.  Finally, inertial and overdamped
stochastic models of the form (\ref{fluct1}) and (\ref{fluct18}) where
the particles interact via a {\it binary} potential (instead of a field equation (\ref{fluct2}) depending on the history of the system) have been studied in
\cite{hb}. A hierarchy of equations has been derived for the reduced probability distributions and Eqs. (\ref{fluct15}) and (\ref{intro1}) are obtained in
a mean field approximation valid in a proper thermodynamic limit with
$N\rightarrow +\infty$.

Let us finally consider the deterministic case without diffusion
$D=T_{eff}=0$. We introduce the exact distribution function and the exact density
\begin{eqnarray}
\label{fluct20}
f_{ex}({\bf r},{\bf v},t)=\sum_{\alpha=1}^{N}\delta ({\bf r}-{\bf r}_{\alpha}(t))\delta ({\bf v}-{\bf v}_{\alpha}(t)),
\end{eqnarray} 
\begin{eqnarray}
\label{fluct21}
\rho_{ex}({\bf r},t)=\sum_{\alpha=1}^{N}\delta ({\bf r}-{\bf r}_{\alpha}(t))=\int f_{ex}({\bf r},{\bf v},t) d{\bf v}.
\end{eqnarray} 
Then, from the equations of motion (\ref{fluct1}), repeating the steps (\ref{fluct7})-(\ref{fluct9}) without the brackets, we  obtain the exact equations
\begin{eqnarray}
\label{fluct22}{\partial f_{ex}\over\partial t}+{\bf v}\cdot {\partial f_{ex}\over\partial {\bf r}}+\nabla c_{ex} \cdot {\partial f_{ex}\over\partial {\bf v}}
={\partial\over\partial {\bf v}}\cdot \left (\xi f_{ex}{\bf v}\right ),  \qquad\qquad  
\end{eqnarray} 
\begin{equation}
\label{fluct23}{\partial c_{ex}\over\partial t}=-kc_{ex}+D_{c}\Delta c_{ex}+h\rho_{ex}.
\end{equation}
These equations bear exactly the same information as the $N$-body
system (\ref{fluct1}) with $D=0$. For $\xi=0$, Eq. (\ref{fluct22}) becomes
equivalent to the Klimontovich equation in plasma physics \cite{pitaevskii}. On the
other hand, in the strong friction limit, if we start from the
$N$-body system (\ref{fluct18}) with $D_{*}=T_{eff}=0$, we obtain the exact equations
\begin{equation}
{\partial\rho_{ex}\over\partial t}=-\chi\nabla\cdot
(\rho_{ex}\nabla c_{ex}),
\label{fluct24}
\end{equation}
\begin{equation}
\label{fluct25}{\partial c_{ex}\over\partial t}=-kc_{ex}+D_{c}\Delta c_{ex}+h\rho_{ex}.
\end{equation}
These equations bear exactly the same information as the $N$-body system (\ref{fluct18}) with $D_{*}=0$.

\section{The limit of strong friction}
\label{sec_sf}

The generalized Smoluchowski equation (\ref{j13}) can be derived from the
generalized Kramers equation (\ref{fp2}) in the strong friction limit
$\xi\rightarrow +\infty$. Using the effective Einstein relation $\xi=D\beta$, the generalized Kramers equation (\ref{fp2}) can be rewritten
\begin{equation}
\label{sf1} {\partial f\over\partial t}+{\bf v}\cdot {\partial
f\over\partial {\bf r}}+\nabla S(c)\cdot {\partial f\over\partial {\bf
v}} ={\partial\over\partial {\bf v}}\cdot \biggl\lbrace D \biggl\lbrack
fC''(f){\partial f\over\partial {\bf v}}+\beta f{\bf
v}\biggr\rbrack\biggr\rbrace.
\end{equation}
In the strong friction limit $\xi=D\beta\rightarrow +\infty$, assuming
$\beta$ of order unity, the term in bracket in Eq. (\ref{sf1}) must
vanish (since $D\rightarrow +\infty$) so that the distribution
function satisfies to leading order
\begin{equation}
\label{sf2}
C'(f)=-\beta\left\lbrack \frac{v^{2}}{2}+\lambda({\bf r},t)\right \rbrack +O(\xi^{-1}).
\end{equation}
The function $\lambda({\bf r},t)$ is related to the spatial density $\rho({\bf r},t)$ through the relation
\begin{equation}
\label{sf3}
\rho({\bf r},t)=\int f d{\bf v}=\rho[\lambda({\bf r},t)].
\end{equation}
Using Eq. (\ref{sf2}), we find  that ${\bf u}=O(\xi^{-1})$ and $P_{ij}=p\delta_{ij}+O(\xi^{-1})$ where $p({\bf r},t)$ is the local pressure 
\begin{equation}
\label{sf4}
p({\bf r},t)=\frac{1}{d}\int f v^{2}d{\bf v}=p[\lambda({\bf r},t)],
\end{equation}
determined from Eqs. (\ref{sf2}) and (\ref{sf3}). As in
Sec. \ref{sec_j}, the fluid is barotropic, i.e.  $p({\bf
r},t)=p\lbrack \rho({\bf r},t)\rbrack$, where the equation of state
$p(\rho)$ is obtained by eliminating $\lambda({\bf r},t)$ between
Eqs. (\ref{sf3}) and (\ref{sf4}). The equation of state is entirely
specified by the function $C(f)$. To first order in $\xi^{-1}$, the momentum
equation (\ref{j3}) implies that
\begin{equation}
\label{sf5} \rho {\bf u}=-{1\over\xi}(\nabla
p-\rho S'(c)\nabla c)+O(\xi^{-2}).
\end{equation}
Inserting the relation (\ref{sf5}) in the continuity equation
(\ref{j2}), we get the generalized Smoluchowski equation
(\ref{j13}). The generalized Smoluchowski equation, as well as the
first order correction to the distribution function $f({\bf r},{\bf
v},t)$, can also be obtained from a formal Chapman-Enskog expansion
\cite{lemou}. Note finally that, instead of the generalized Fokker-Planck equation (\ref{sf1}), we could consider the generalized {\it isotropic} BGK
equation \cite{lemou}:
\begin{equation}
\label{sf6} {\partial f\over\partial t}+{\bf v}\cdot {\partial
f\over\partial {\bf r}}+\nabla S(c)\cdot {\partial f\over\partial {\bf
v}} =\frac{f-f_{0}}{\tau},
\end{equation}
where $f_{0}({\bf r},{\bf v},t)$ is defined by
$C'(f_{0})=-\beta({v^{2}}/{2}+\lambda({\bf r},t))$ where $\lambda({\bf r},t)$ is determined by $\int
f_{0}d{\bf v}=\rho({\bf r},t)$. The calculations are similar to those
described previously with $1/\tau$ playing the role of $\xi$. In
particular, the generalized Smoluchowski equation (\ref{j13}) is
obtained at order $O(\tau)$ for $\tau\rightarrow 0$. A Chapman-Enskog 
expansion is given in Appendix A of \cite{lemou}.

\section{The limit of large diffusivity}
\label{sec_ld}

Let us consider the Keller-Segel model 
\begin{equation}
{\partial\rho\over\partial t}=D_*\Delta\rho-\chi\nabla\cdot
(\rho\nabla c), \label{ld1}
\end{equation}
\begin{equation}
\label{ld2}{\partial c\over\partial t}=-k c+h \rho+D_{c}\Delta
c,
\end{equation}
with Neumann boundary conditions 
\begin{equation}
\nabla\rho\cdot {\bf n}=\nabla c\cdot {\bf n}=0,\label{ld2b}
\end{equation} 
where ${\bf n}$ is a unit vector normal to the boundary of the
domain. Following J\"ager \& Luckhaus \cite{jl}, we set $h=\lambda
D_{c}$ and assume that $D_{c}\rightarrow +\infty$ and $\lambda\sim
1$. Introducing the average density $\overline{\rho}=(1/V)\int \rho \,
d{\bf r}$ (where $V$ is the volume of the box), we find from
Eq. (\ref{ld1}) that $\overline{\rho}(t)=\overline{\rho}_{0}$. Then,
from Eq. (\ref{ld2}), we get
\begin{equation}
\label{ld3}
{1\over D_{c}}\left ({\partial \overline{c}\over\partial t}+k\overline{c}\right )=\lambda\overline{\rho}_{0}.
\end{equation}
We note, parenthetically, that this equation has the exact solution,
\begin{equation}
\label{ld3bis}
\overline{c}(t)=\frac{h}{k}\overline{\rho}_{0}(1-e^{-kt})+\overline{c}_{0}e^{-kt},
\end{equation}
so that the average concentration of the chemical relaxes to $\overline{c}(+\infty)=(h/k)\overline{\rho}_{0}$ on a timescale $k^{-1}$. If we consider $\tilde c=c-\overline{c}$, we find from Eqs. (\ref{ld2}) and (\ref{ld3}) that
\begin{equation}
\label{ld4}
{1\over D_{c}}\left ({\partial \tilde{c}\over\partial t}+k\tilde{c}\right )=\Delta\tilde c+\lambda (\rho-\overline{\rho}_{0}).
\end{equation}
In the limit $D_{c}\rightarrow +\infty$, we obtain the reduced Keller-Segel model 
\begin{equation}
{\partial\rho\over\partial t}=D_*\Delta\rho-\chi\nabla\cdot
(\rho\nabla c), \label{ld5}
\end{equation}
\begin{equation}
\label{ld6}\Delta c=-\lambda(\rho-\overline{\rho}_{0}).
\end{equation}
Note that this system is mathematically well-posed with the Neumann boundary conditions (\ref{ld2b}), contrary to the case where Eq. (\ref{ld6}) is replaced by the Poisson equation
\begin{equation}
\label{ld7}
\Delta c=-\lambda\rho.
\end{equation} 
Indeed, if we integrate Eq. (\ref{ld7}) over the domain and use the divergence theorem, we obtain a contradiction as
\begin{equation}
\label{ld7b}
\oint \nabla c \cdot d{\bf S}=\int \Delta c \ d{\bf r}=  -\lambda M\neq 0.
\end{equation} 
However, when the density blows up so that $\rho\gg
\overline{\rho}_{0}$, it is justified to consider the model
(\ref{ld5})-(\ref{ld7}) as an approximation. {\it It is only in that
case (large diffusivity of the chemical and high concentration of the
particles) that the Keller-Segel model for the chemotaxis becomes
equivalent to the Smoluchowski-Poisson system for self-gravitating
Brownian particles.}  On the other hand, we note that a homogeneous
distribution is an exact stationary solution of the Keller-Segel model
(\ref{ld1})-(\ref{ld2}) with $k c=h \rho$. It is also an exact
stationary solution of the reduced Keller-Segel model
(\ref{ld5})-(\ref{ld6}) with $\rho=\overline{\rho}_{0}$. By contrast,
a homogeneous distribution is {\it not} a stationary solution of
Eqs. (\ref{ld5}) and (\ref{ld7}) when Eq. (\ref{ld6}) is replaced by a
Poisson equation (\ref{ld7}) as in astrophysics. In astrophysics, we
have to advocate the Jeans swindle when we analyse the linear
dynamical stability of an infinite and homogeneous self-gravitating
system
\cite{bt}. By contrast, {\it there is no ``Jeans swindle'' in the
chemotactic problem} if we properly use Eqs. (\ref{ld2}) or
(\ref{ld6}) instead of Eq. (\ref{ld7}) \cite{epjbjeans}.

Setting $k=k_{0}^{2} D_{c}$ and $h=\lambda D_{c}$, we can also consider the
limit $D_{c}\rightarrow +\infty$ with $\lambda, k_{0} \sim 1$. Equation
(\ref{ld2}) can be rewritten
\begin{equation}
\label{ld8}
\frac{1}{D_{c}}{\partial c\over\partial t}=-k_{0}^{2} c+\lambda \rho+\Delta c,
\end{equation}
which reduces, for $D_{c}\rightarrow
+\infty$, to
\begin{equation}
\label{ld9}
\Delta c-k_{0}^{2} c=-\lambda \rho.
\end{equation}

\section{Kinetic derivation of the hyperbolic model}
\label{sec_hyp}

The kinetic theory presented in Sec. \ref{sec_stoc} is well-suited to
{\it weakly inertial} systems for which the damping term $-\xi {\bf u}$ is
relatively strong. For $\xi\rightarrow +\infty$, we rigorously obtain
the generalized Smoluchowski equation (\ref{j13}). For large values of
$\xi$, the damped Euler equations (\ref{j9})-(\ref{j11}) may provide a
relatively good description of the dynamics (but we again stress that
they are not rigorously justified). Alternatively, the model
(\ref{intro3})-(\ref{intro5}) proposed by Gamba {\it et al.}
\cite{gamba} corresponds to {\it highly inertial} systems. Formally, it can
be obtained from Eqs. (\ref{j9})-(\ref{j11}) by taking $-\xi {\bf
u}={\bf 0}$. However, we have indicated in Sec. \ref{sec_j} the
shortcoming of this procedure when we start from a stochastic equation
of the form (\ref{fluct1}). Indeed, the L.T.E. condition (\ref{j6}) is
not justified when $\xi\rightarrow 0$. Here, we present an alternative
kinetic model leading rigorously to the model
(\ref{intro3})-(\ref{intro5}) proposed by Gamba {\it et al.}
\cite{gamba}. We assume that the particles obey a stochastic equation
of the form
\begin{equation}
\label{hyp1} {d{\bf v}\over dt}=-\zeta ({\bf v}-{\bf u}({\bf r},t))+\nabla S(c)+\sqrt{2D(f)}{\bf
R}(t).
\end{equation}
The first term in the r.h.s. is a friction force {\it relative} to the
mean velocity ${\bf u}({\bf r},t)$. Its physical effect is therefore
completely different from the friction force in Eq. (\ref{lang1}). The
associated nonlinear mean field Fokker-Planck equation is
\begin{equation}
\label{hyp2} {\partial f\over\partial t}+{\bf v}\cdot {\partial
f\over\partial {\bf r}}+\nabla S(c)\cdot {\partial f\over\partial {\bf
v}} ={\partial\over\partial {\bf v}}\cdot \biggl\lbrace D \biggl\lbrack
fC''(f){\partial f\over\partial {\bf v}}+\beta f({\bf
v}-{\bf u}({\bf r},t))\biggr\rbrack\biggr\rbrace,
\end{equation}
where $\beta=\zeta/D$.  Taking the hydrodynamical moments of this
equation, as in Sec. \ref{sec_j}, we obtain the continuity equation
(\ref{j2}) and
\begin{equation}
\label{hyp3} {\partial\over\partial t}(\rho
u_{i})+{\partial\over\partial x_{j}}(\rho u_{i}u_{j})= -{\partial
P_{ij}\over\partial x_{j}}+\rho S'(c){\partial c\over\partial
x_{i}}.
\end{equation}
Contrary to Eq. (\ref{j3}), this equation has no damping term and it
conserves the total impulse. Considering the limit
$\zeta=D\beta\rightarrow +\infty$ with $\beta$ fixed, we find from
Eq. (\ref{hyp2}) that the distribution function satisfies
\begin{equation}
\label{hyp4} C'(f)=-\beta\biggl \lbrack {w^{2}\over
2}+\lambda({\bf r},t)\biggr \rbrack+O(\zeta^{-1}).
\end{equation}
Therefore, in this approach, the L.T.E. (\ref{j6}) is exact to leading
order in $\zeta\rightarrow +\infty$. This implies that
$P_{ij}=p\delta_{ij}+O(\zeta^{-1})$ where the pressure $p({\bf r},t)$
satisfies a barotropic equation of state $p(\rho)$ determined by the
function $C(f)$ as in Sec. \ref{sec_j}. Substituting this result in
Eq. (\ref{hyp3}), we get the hydrodynamic model
(\ref{intro3})-(\ref{intro5}) without friction term proposed by Gamba
{\it et al.} \cite{gamba}. Note finally that, instead of the
generalized Fokker-Planck equation (\ref{hyp2}) we could consider the
generalized BGK equation
\begin{equation}
\label{hyp5} {\partial f\over\partial t}+{\bf v}\cdot {\partial
f\over\partial {\bf r}}+\nabla S(c)\cdot {\partial f\over\partial {\bf
v}} =\frac{f-f_{L.T.E.}}{\tau},
\end{equation}
where $f_{L.T.E.}({\bf r},{\bf v},t)$ is defined by Eq. (\ref{j6}).
The calculations are similar to those developed previously with
$1/\tau$ playing the role of $\zeta$. In particular, the model
(\ref{intro3})-(\ref{intro5}) is obtained at order $O(\tau)$ for
$\tau\rightarrow 0$. Note, finally, that other kinetic theories for
chemosensitive movement have been proposed in \cite{filbet}.  They lead
to hydrodynamic equations of the form (\ref{intro3})-(\ref{intro5}),
without friction force.

\section{Generalized free energies and Lyapunov functionals}
\label{sec_fe}

In this Appendix, we show that the different equations introduced in this
paper are associated with Lyapunov functionals that can be interpreted
as effective ``generalized free energies''.

\subsection{Kinetic model}
\label{sec_km}

Let us first consider the kinetic model (\ref{fp2})-(\ref{lang2}) with
constant coefficients
\begin{equation}
\label{fe0} {\partial f\over\partial t}+{\bf v}\cdot {\partial
f\over\partial {\bf r}}+\nabla c\cdot {\partial f\over\partial {\bf
v}} ={\partial\over\partial {\bf v}}\cdot \biggl\lbrack
D f C''(f){\partial f\over\partial {\bf v}}+\xi f{\bf
v}\biggr\rbrack,
\end{equation}
\begin{equation}
\label{fe1}{\partial c\over\partial t}=-k c+h \rho+D_{c}\Delta
c.
\end{equation}
Recalling that $T_{eff}=D/\xi$, we introduce the functional
\begin{eqnarray}
\label{fe2}
F[f]=\int f\frac{v^{2}}{2}d{\bf r}d{\bf v}+\frac{1}{2h}\int \left\lbrack D_{c}(\nabla c)^{2}+kc^{2}\right\rbrack d{\bf r}
-\int \rho c d{\bf r}+T_{eff}\int C(f)d{\bf r}d{\bf v}.
\end{eqnarray}
After simple algebra, one can  show that
\begin{eqnarray}
\label{fe3}
\dot F=-\int \frac{1}{\xi f} \left (D f C''(f) \frac{\partial f}{\partial {\bf v}}+\xi f {\bf v}\right )^{2} d{\bf r}d{\bf v}
-\frac{1}{h}\int (D_{c}\Delta c-kc+h\rho)^{2} d{\bf r} \le 0.
\end{eqnarray}
Therefore, $F(t)$ decreases monotonically with time and plays the role of a
Lyapunov functional. It can also be interpreted as a generalized free
energy \cite{gen}. This is particularly clear if we consider the field
equation 
\begin{equation}
\label{fe4}D_{c}\Delta c-k c=-h \rho,
\end{equation}
instead of Eq. (\ref{fe1}) [see Appendix \ref{sec_ld}]. In that case, the functional (\ref{fe2}) reduces to
\begin{eqnarray}
\label{fe5}
F=\int f\frac{v^{2}}{2}d{\bf r}d{\bf v}
-\frac{1}{2}\int \rho c d{\bf r}+T_{eff}\int C(f)d{\bf r}d{\bf v}.
\end{eqnarray}
This can be written $F=E-T_{eff}S$ where $E=\int f\frac{v^{2}}{2}d{\bf
r}d{\bf v} -\frac{1}{2}\int \rho c d{\bf r}=K+W$ represents the energy
(kinetic $+$ potential) and $S=-\int C(f)d{\bf r}d{\bf v}$ a
generalized entropy. In that case, we have
\begin{eqnarray}
\label{fe7}
\dot F=-\int \frac{1}{\xi f} \left (D f C''(f) \frac{\partial f}{\partial {\bf v}}+\xi f {\bf v}\right )^{2} d{\bf r}d{\bf v} \le 0.
\end{eqnarray}
This inequality can be viewed as an appropriate $H$-theorem in the
canonical ensemble \cite{gen,hb}, where the effective temperature
$T_{eff}$ is fixed instead of the energy. In the case of normal
diffusion $C(f)=f\ln f$, the functional (\ref{fe5}) coincides with the
Boltzmann free energy
\begin{eqnarray}
\label{fe5bb}
F=\int f\frac{v^{2}}{2}d{\bf r}d{\bf v}
-\frac{1}{2}\int \rho c d{\bf r}+T_{eff}\int f \ln f d{\bf r}d{\bf v}.
\end{eqnarray}
We also note, parenthetically, that in the
absence of diffusion ($D=T_{eff}=0$), the free energy reduces to the
energy. Therefore, the proper $H$-theorem becomes  $\dot E\le 0$. More
precisely, we have $\dot E=-\xi \int f v^{2}d{\bf r}d{\bf v}=-2\xi K\le 0$,
where $K$ is the kinetic energy. 

The steady state of the system (\ref{fe0})-(\ref{fe1}) must satisfy $\dot F=0$. According to Eq. (\ref{fe3}), this implies that the current in Eq. (\ref{fe0}) vanishes
\begin{equation}
\label{fe8}
C''(f){\partial f\over\partial {\bf v}}+\beta {\bf
v}={\bf 0}.
\end{equation}
Then, the condition $\partial f/\partial t=0$ implies that the advective term must also vanish 
\begin{equation}
\label{fe9} {\bf v}\cdot {\partial
f\over\partial {\bf r}}+\nabla c\cdot {\partial f\over\partial {\bf
v}} =0.
\end{equation}
Therefore, the advective term and the current vanish {\it independently}.
Equation (\ref{fe8}) can be integrated on the velocity to yield
\begin{equation}
\label{fe10}
C'(f)=-\beta \frac{v^{2}}{2}-A({\bf r}),
\end{equation}
where $A({\bf r})$ is an arbitrary function of the position.
Differentiating this expression with respect to ${\bf r}$ and ${\bf
v}$, we obtain
\begin{equation}
\label{fe11}
C''(f)\frac{\partial f}{\partial {\bf r}}=-\nabla A({\bf r}), \qquad C''(f)\frac{\partial f}{\partial {\bf v}}=-\beta {\bf v}.
\end{equation}
Substituting these relations in Eq. (\ref{fe9}), we get
\begin{equation}
\label{fe13}
(\nabla A+\beta \nabla c)\cdot {\bf v}=0,
\end{equation}
which must be valid for all ${\bf v}$. This yields $\nabla A+\beta \nabla c={\bf 0}$, so that
\begin{equation}
\label{fe14}
A({\bf r})=-\beta c({\bf r})+\alpha, 
\end{equation}
where $\alpha$ is a constant of integration. Substituting this result
in Eq. (\ref{fe10}), we obtain the steady state (\ref{fp3}). Therefore, at
equilibrium, the distribution function depends only on the energy:
$f=f(\epsilon)$ with $\epsilon=v^{2}/2-c({\bf r})$. This first implies
that the local velocity ${\bf u}({\bf r})={\bf 0}$. The density and
the pressure can then be written $\rho=\int f(\epsilon)d{\bf v}$ and
$p=\frac{1}{d}\int f(\epsilon) v^{2}d{\bf v}$. Since
\begin{equation}
\label{fe14bgt}
\nabla p=-\frac{1}{d}\nabla c\int f'(\epsilon)v^{2}d{\bf v}=-\frac{1}{d}\nabla c\int \frac{\partial f}{\partial {\bf v}}\cdot {\bf v}d{\bf v}=\nabla c\int f d{\bf v}=\rho\nabla c, 
\end{equation}
we find that the condition $f=f(\epsilon)$ implies the condition of
hydrostatic equilibrium
\begin{eqnarray}
\label{fe25}
\nabla p-\rho\nabla c={\bf 0}.
\end{eqnarray}

We can also justify the Local Thermodynamical Equilibrium (LTE)
assumption made in Sec. \ref{sec_j} from the free energy (\ref{fe2}).
The idea is to close the hierarchy of hydrodynamic equations by
calculating the pressure tensor (\ref{j4}) with the distribution
$f_{LTE}$ that minimizes the free energy (\ref{fe2}) at fixed
$\rho({\bf r},t)$ and ${\bf u}({\bf r},t)$.  If we prescribe the
density $\rho({\bf r},t)$, the concentration of the chemical $c({\bf r},t)$ is
automatically fixed by Eq. (\ref{fe1}). This implies that the second and third integrals in Eq. (\ref{fe2}) are fixed. Therefore, minimizing $F$ at
fixed $\rho({\bf r},t)$ and ${\bf u}({\bf r},t)$ is equivalent to
minimizing
\begin{eqnarray}
\label{fe15}
\tilde F=\int f\frac{v^{2}}{2}d{\bf r}d{\bf v}+T_{eff}\int C(f)\, d{\bf r}d{\bf v},
\end{eqnarray}
at fixed $\rho({\bf r},t)$ and ${\bf u}({\bf r},t)$. Introducing Lagrange multipliers and writing the variational problem in the form
\begin{eqnarray}
\label{fe16}
\delta F-\int a({\bf r},t)\delta f \, d{\bf r}d{\bf v}-\int  {\bf b}({\bf r},t)\delta f \cdot {\bf v}\, d{\bf r}d{\bf v}=0,
\end{eqnarray}
we obtain
\begin{eqnarray}
\label{fe17}
\frac{v^{2}}{2}+T_{eff}C'(f)-a({\bf r},t)-{\bf b}({\bf r},t)\cdot {\bf v}=0.
\end{eqnarray}
Relating the Lagrange multipliers $a({\bf r},t)$ and ${\bf b}({\bf
r},t)$ to the constraints $\rho({\bf r},t)$ and ${\bf u}({\bf r},t)$,
we obtain the distribution (\ref{j6}). Since $\delta^{2}\tilde
F=(1/2)T_{eff}\int C''(f)(\delta f)^{2}d{\bf r}d{\bf v}\ge 0$,
and $C$ is convex, the distribution (\ref{j6}) is a minimum of
$F$ at fixed $\rho({\bf r},t)$ and ${\bf u}({\bf r},t)$.

\subsection{Fluid model}

Let us consider the damped barotropic Euler equations with constant
coefficients
\begin{equation}
\label{fe18} {\partial\rho\over\partial t}+\nabla\cdot (\rho{\bf
u})=0,
\end{equation}
\begin{equation}
\label{fe19} {\partial {\bf u}\over\partial t}+({\bf u}\cdot
\nabla){\bf u}=-{1\over\rho}\nabla p+\nabla c-\xi {\bf u},
\end{equation}
\begin{equation}
\label{fe20}{\partial c\over\partial t}=-k c+h \rho+D_{c}\Delta c.
\end{equation}
We introduce the functional
\begin{eqnarray}
\label{fe21}
F[\rho,{\bf u}]=\int \rho\int^{\rho}\frac{p(\rho')}{\rho'^{2}}d\rho' d{\bf r}+\frac{1}{2h}\int \left\lbrack D_{c}(\nabla c)^{2}+kc^{2}\right\rbrack d{\bf r}
-\int \rho c d{\bf r}+\frac{1}{2}\int \rho {\bf u}^{2}d{\bf r}.
\end{eqnarray}
This functional can be obtained from Eq. (\ref{fe2}) by using the LTE
condition (\ref{j6}) to express $F[f]$ as a functional of $\rho$ and ${\bf u}$
(the calculations are similar to those detailed in \cite{cban,lemou}). After simple algebra, it is
easy to show that
\begin{eqnarray}
\label{fe22}
\dot F=-\xi \int \rho {\bf u}^{2} d{\bf r}
-\frac{1}{h}\int (D_{c}\Delta c-kc+h\rho)^{2} d{\bf r} \le 0.
\end{eqnarray}
Therefore, $F(t)$ decreases monotonically with time and plays the role
of a Lyapunov functional. It can also be interpreted as a generalized
free energy. In the case where Eq. (\ref{fe20}) is replaced by
Eq. (\ref{fe4}), we have
\begin{eqnarray}
\label{fe23}
F[\rho,{\bf u}]=\int \rho\int^{\rho}\frac{p(\rho')}{\rho'^{2}}d\rho' d{\bf r}
-\frac{1}{2}\int \rho c \, d{\bf r}+\frac{1}{2}\int \rho {\bf u}^{2}d{\bf r}.
\end{eqnarray}
and
\begin{eqnarray}
\label{fe24}
\dot F=-\xi \int \rho {\bf u}^{2} d{\bf r}
 \le 0.
\end{eqnarray}
Note that for an isothermal equation of state $p=\rho T_{eff}$, corresponding to normal diffusion, the functional (\ref{fe23}) returns the Boltzmann free energy
\begin{eqnarray}
\label{fe23bb}
F[\rho,{\bf u}]=T_{eff}\int \rho\ln\rho d{\bf r}
-\frac{1}{2}\int \rho c d{\bf r}+\frac{1}{2}\int \rho {\bf u}^{2}d{\bf r}.
\end{eqnarray}
The steady states of Eqs. (\ref{fe18})-(\ref{fe20}) must satisfy $\dot F=0$ yielding, according to Eq. (\ref{fe22}), ${\bf u}={\bf 0}$. Then, using Eqs. (\ref{fe18})-(\ref{fe20}), we obtain the 
condition of hydrostatic equilibrium (\ref{fe25}).

\subsection{Overdamped model}

Let us consider the generalized Smoluchowski equation with constant
coefficients
\begin{equation}
{\partial\rho\over\partial t}=\nabla\cdot \biggl \lbrack
\frac{1}{\xi}\left (\nabla p-\rho \nabla c\right
)\biggr\rbrack, \label{fe26}
\end{equation}
\begin{equation}
\label{fe27}{\partial c\over\partial t}=-k c+h \rho+D_{c}\Delta c.
\end{equation}
We introduce the functional
\begin{eqnarray}
\label{fe28}
F[\rho]=\int \rho\int^{\rho}\frac{p(\rho')}{\rho'^{2}}d\rho' d{\bf r}+\frac{1}{2h}\int \left\lbrack D_{c}(\nabla c)^{2}+kc^{2}\right\rbrack d{\bf r}
-\int \rho c \, d{\bf r}.
\end{eqnarray}
This functional can be obtained from Eq. (\ref{fe2}) by using
Eq. (\ref{sf2}) to express $F[f]$ as a functional of $\rho$ in the
strong friction limit \cite{cban,lemou}. It can also be obtained from
Eq. (\ref{fe21}) by using the fact that ${\bf u}=O(\xi^{-1})$. After
simple algebra, it is easy to show that
\begin{eqnarray}
\label{fe29}
\dot F=-\int \frac{1}{\xi\rho}(\nabla p-\rho\nabla c)^{2}d{\bf r}
-\frac{1}{h}\int (D_{c}\Delta c-kc+h\rho)^{2} d{\bf r} \le 0.
\end{eqnarray}
Therefore, $F(t)$ decreases monotonically with time and plays the role
of a Lyapunov functional. It can also be interpreted as a generalized
free energy. In the case where Eq. (\ref{fe27}) is replaced by
Eq. (\ref{fe4}), we have
\begin{eqnarray}
\label{fe30}
F[\rho]=\int \rho\int^{\rho}\frac{p(\rho')}{\rho'^{2}}d\rho' d{\bf r}
-\frac{1}{2}\int \rho c d{\bf r},
\end{eqnarray}
and
\begin{eqnarray}
\label{fe31}
\dot F=-\int \frac{1}{\xi\rho}(\nabla p-\rho\nabla c)^{2}d{\bf r}\le 0.
\end{eqnarray}
Note that for an isothermal equation of state, $p=\rho T_{eff}$, we obtain the Boltzmann free energy in configuration space
\begin{eqnarray}
\label{fe30bb}
F[\rho]=T_{eff}\int \rho\ln\rho \, d{\bf r}
-\frac{1}{2}\int \rho c \, d{\bf r}.
\end{eqnarray}
The steady states of Eqs. (\ref{fe26})-(\ref{fe27}) must satisfy $\dot
F=0$ yielding, according to Eq. (\ref{fe29}), the condition of
hydrostatic equilibrium (\ref{fe25}).

\subsection{The primitive Keller-Segel model}
\label{sec_prim}

The primitive Keller-Segel model has the form \cite{ks}:
\begin{eqnarray}
\label{fe33}
\frac{\partial\rho}{\partial t}=\nabla\cdot (D_{2}\nabla\rho)-\nabla\cdot (D_{1}\nabla c),
\end{eqnarray}
\begin{equation}
\label{fe34}{\partial c\over\partial t}=-k(c) c+h(c) \rho+D_{c}\Delta c,
\end{equation}
where $D_{1}=D_{1}(\rho,c)$ and $D_{2}=D_{2}(\rho,c)$ can both depend
on the concentration of cells and of the chemical. Let us consider a simplification where $D_{2}=D h(\rho)$, $D_{1}=\chi g(\rho)$, $k(c)=k$ and  $h(c)=h$. In that case, we obtain 
\begin{eqnarray}
\label{fe35}
\frac{\partial\rho}{\partial t}=\nabla\cdot \left\lbrack Dh(\rho)\nabla\rho-\chi g(\rho)\nabla c\right\rbrack,
\end{eqnarray}
\begin{equation}
\label{fe36}{\partial c\over\partial t}=-k c+h \rho+D_{c}\Delta c.
\end{equation}
Equation (\ref{fe35}) can be viewed as a nonlinear Fokker-Planck
equation of the form considered in \cite{gen}. It can be obtained from 
the stochastic process
\begin{eqnarray}
\label{fe35bis}
\frac{d{\bf r}}{dt}=\chi(\rho)\nabla c+\sqrt{2D(\rho)}{\bf R}(t),
\end{eqnarray}
where ${\bf R}(t)$ is a white noise, $\chi(\rho)=\chi g(\rho)/\rho$
and $D(\rho)=(D/\rho)\int^{\rho}h(\rho')d\rho'$. When $g(\rho)=\rho$
and $h(\rho)=1$, we recover the ordinary Keller-Segel model
(\ref{intro1}) with constant diffusion coefficient $D$ and constant
mobility $\chi$. When $g(\rho)=\rho$ and $h(\rho)$ is arbitrary,
Eq. (\ref{fe35}) describes a situation where the mobility is constant
but the diffusion coefficient can depend on the density. This can
account for anomalous diffusion and non-ergodic effects. This is the
case in Eqs. (\ref{gsk13}) and (\ref{gsk16}) corresponding to
$g(\rho)=\rho$ and $h(\rho)=(\chi/D)K\gamma\rho^{\gamma-1}$
\cite{lang} or, more generally, in Eq. (\ref{add1}) corresponding to
$g(\rho)=\rho$ and $h(\rho)=(\chi/D)p'(\rho)$ \cite{gen}.  When
$h(\rho)=1$ and $g(\rho)$ is nonlinear, Eq. (\ref{fe35}) describes a
situation where the diffusion coefficient is constant but the mobility
(or chemotactic sensitivity) depends on the density. This can account
for excluded volume effects or steric hindrance when the density is
high. For example, the case where $h(\rho)=1$ and $g(\rho)=\rho
(1-\rho/\sigma_{0})$ has been treated in \cite{hillen,degrad}. In that
case, the mobility tends to zero when the density approaches its
maximum value $\sigma_{0}$. More generally, the drift-diffusion
equation (\ref{fe35}) describes a situation where the diffusion
coefficient $D(\rho)$ and the mobility $\chi(\rho)$ can both depend on
the density. It can be derived from a master equation by allowing the
transition probabilities from one site to the other to depend on the
occupancy number
\cite{kaniadakis,degrad}.

 If we  introduce the functional
\begin{eqnarray}
\label{fe37}
F[\rho]=\frac{D}{\chi}\int C(\rho) \, d{\bf r}+\frac{1}{2h}\int \left\lbrack D_{c}(\nabla c)^{2}+kc^{2}\right\rbrack d{\bf r}
-\int \rho c \, d{\bf r},
\end{eqnarray}
where $C(\rho)$ satisfies $C''(\rho)={h(\rho)}/{g(\rho)}$, it is easy to show after simple algebra that
\begin{eqnarray}
\label{fe38}
\dot F=-\int \frac{1}{\chi g(\rho)}(D h(\rho)\nabla\rho-\chi g(\rho)\nabla c)^{2}d{\bf r}
-\frac{1}{h}\int (D_{c}\Delta c-kc+h\rho)^{2} d{\bf r} \le 0.
\end{eqnarray}
Therefore, $F(t)$ decreases monotonically with time and plays the role
of a Lyapunov functional. In the case where Eq. (\ref{fe36}) is replaced by
Eq. (\ref{fe4}), we have
\begin{eqnarray}
\label{fe39}
F[\rho]=\frac{D}{\chi}\int C(\rho) \, d{\bf r}
-\frac{1}{2}\int \rho c \, d{\bf r},
\end{eqnarray}
and
\begin{eqnarray}
\label{fe40}
\dot F=-\int \frac{1}{\chi g(\rho)}(D h(\rho)\nabla\rho-\chi g(\rho)\nabla c)^{2}d{\bf r}\le 0.
\end{eqnarray}
This can be written $F=E-T_{eff}S$ where $E=-(1/2)\int \rho c
d{\bf r}$ is the potential energy, $T_{eff}=1/\beta$ is an effective
inverse temperature satisfying a generalized Einstein relation
$\beta=\chi/D$ and $S=-\int C(\rho) d{\bf r}$ is a generalized
entropy. Therefore, $F$ can be interpreted as a generalized free
energy in an effective thermodynamical formalism. We can thus write
Eq.  (\ref{fe35}) in different forms as shown explicitly in
\cite{gen,degrad}. This  strengthen the analogy with generalized Fokker-Planck equations.  The steady states of Eqs. (\ref{fe35})-(\ref{fe36}) must
satisfy $\dot F=0$.  According to Eq. (\ref{fe38}), this implies that the current must vanish yielding the
relation 
\begin{eqnarray}
\label{fe40bis}
C'(\rho_{eq})=\beta c-\alpha,
\end{eqnarray}
where $\alpha$ is a constant of integration.  Since $C$ is convex, the
above relation can be reversed. Then, we find that $\rho_{eq}({\bf
r})=F(-\beta c+\alpha)=\rho_{eq}(c)$ where $F(x)=(C')^{-1}(-x)$. Since
$\rho_{eq}'(c)=\beta/C''(\rho_{eq})$ and $\beta>0$, we find that
$\rho_{eq}(c)$ is a monotonically increasing function of the
concentration.  Equation (\ref{fe40bis}) can also be obtained by
extremizing the free energy (\ref{fe37}) at fixed mass. Furthermore,
it is shown in
\cite{gen} that linearly dynamically stable solutions of
(\ref{fe35})-(\ref{fe36}) correspond to {\it minima} of $F$ at fixed
mass (these general results also apply to the other model equations
discussed previously).  Finally, we note that Eq. (\ref{fe35}) can be
written in the form
\begin{eqnarray}
\label{fe42}
\frac{\partial\rho}{\partial t}=\nabla\cdot \left\lbrack \chi g(\rho)\nabla \frac{\delta F}{\delta\rho}\right\rbrack,
\end{eqnarray}
where $\delta/\delta\rho$ denotes the functional derivative \cite{next}.

\section{Generalized Cahn-Hilliard equations}
\label{sec_ch}

In this Appendix, we show that, in the limit of short-range
interactions, the kinetic equations presented in this paper reduce to
generalized forms of the Cahn-Hilliard equation describing phase ordering
kinetics \cite{bray}. The connection to the Cahn-Hilliard equation 
was previously mentioned in \cite{next,lemou}.

Let us first consider the situation where the equation for the chemical
is given by (see Appendix \ref{sec_ld})
\begin{equation}
\label{ch1}\Delta c-k_{0}^{2}c=-\lambda \rho,
\end{equation}
In the limit $k_{0}\rightarrow +\infty$, we can neglect the Laplacian and obtain in first approximation $c\simeq \lambda\rho/k_{0}^{2}$. Then, substituting this relation in the Laplacian we get the next order correction
\begin{equation}
\label{ch2}c\simeq \frac{\lambda}{k_{0}^{4}}(\Delta \rho+k_{0}^{2}\rho).
\end{equation}
More generally, suppose that the concentration of the chemical is determined by a relation of the form
\begin{equation}
\label{ch3}
c({\bf r},t)=\int u(|{\bf r}-{\bf r}'|)\rho({\bf r}',t)d{\bf r}',
\end{equation}
where $u$ is a binary potential of interaction. Let us now assume that
$u(|{\bf r}-{\bf r}'|)$ is a short-range potential of
interaction. Then, setting ${\bf q}={\bf r}'-{\bf r}$ and writing
\begin{equation}
\label{ch4}
c({\bf r},t)=\int u(q)\rho({\bf r}+{\bf q},t)d{\bf q},
\end{equation}
we can Taylor expand $\rho({\bf r}+{\bf q},t)$ to second order in ${\bf q}$
so that
\begin{equation}
\label{ch5}
\rho({\bf r}+{\bf q},t)=\rho({\bf r},t)+\sum_{i}\frac{\partial\rho}{\partial x_{i}}q_{i}+\frac{1}{2}\sum_{i,j}\frac{\partial^{2}\rho}{\partial x_{i}\partial x_{j}}q_{i}q_{j}.
\end{equation} 
Substituting this expansion in Eq. (\ref{ch4}), we obtain
\begin{equation}
\label{ch6}
c({\bf r},t)=a \rho({\bf r},t)+\frac{b}{2}\Delta\rho({\bf r},t), 
\end{equation}
with
\begin{equation}
\label{ch7}
a=S_{d}\int_{0}^{+\infty} u(q) q^{d-1} dq,\qquad b=\frac{1}{d} S_{d}\int_{0}^{+\infty} u(q) q^{d+1} dq.
\end{equation}
Note that $l=(b/a)^{1/2}$ has the dimension of a length. In the limit of short-range interactions, we can replace $c({\bf r},t)$ by Eq. (\ref{ch6}) in all the dynamical models introduced in this paper. Of particular interest is the case of the generalized drift-diffusion equation (\ref{fe35}). Substituting Eq. (\ref{ch6}) in Eq. (\ref{fe35}) and introducing the effective potential
\begin{equation}
\label{ch9}
V(\rho)=-\frac{a}{b}\rho^{2}+\frac{2D}{\chi b} C(\rho)+V_{0},
\end{equation} 
we get
\begin{equation}
\label{ch10}
\frac{\partial\rho}{\partial t}=-A\nabla\cdot \left\lbrack g(\rho)\nabla \left (\Delta\rho-V'(\rho)\right)\right\rbrack,
\end{equation} 
with $A=\chi b/2$. Morphologically, this equation is similar to the
Cahn-Hilliard equation \cite{bray}. One difference, however, is that
$g(\rho)$ can depend on the density (for constant mobility
$g(\rho)=\rho$) while this quantity is constant in the ordinary
Cahn-Hilliard equation. With this (important) difference in mind, the
drift-diffusion equation (\ref{fe35}) can be seen as a generalization
of the Cahn-Hilliard equation to the case of long-range potentials of
interaction. For the particular case $D=T_{eff}=0$, we get
\begin{equation}
\label{ch12}
\frac{\partial\rho}{\partial t}=-A\nabla\cdot \left\lbrack g(\rho)\nabla \left (\Delta\rho+\frac{2a}{b}\rho\right)\right\rbrack.
\end{equation} 
Note, finally, that Eq. (\ref{ch10}) can be written in the form
\begin{eqnarray}
\label{ch13}
\frac{\partial\rho}{\partial t}=A \nabla\cdot \left\lbrack  g(\rho)\nabla \frac{\delta F}{\delta\rho}\right\rbrack,
\end{eqnarray}
where 
\begin{eqnarray}
\label{ch14}
F\lbrack\rho\rbrack=\int \left\lbrack \frac{1}{2} (\nabla\rho)^{2}+V(\rho)\right\rbrack d{\bf r}.
\end{eqnarray}
This expression of the free energy can be obtained from
Eq. (\ref{fe39}), by using Eq. (\ref{ch6}). For comparison, the ordinary Cahn-Hilliard equation for model B (conserved dynamics) is \cite{bray}:
\begin{eqnarray}
\label{ch15}
\frac{\partial\rho}{\partial t}=\Delta \frac{\delta F}{\delta\rho}.
\end{eqnarray}
In the Cahn-Hilliard problem, the potential has a double-well shape leading to a phase separation while, in the present case, the potential can take other forms, as in Eq. (\ref{ch12}) for example. 





\begin{thebibliography}{}





\bibitem{houches}  {\small {\it Dynamics and thermodynamics of systems with long range interactions}, edited by Dauxois, T., Ruffo, S., Arimondo, E. and  Wilkens, M. Lecture Notes in Physics, Springer (2002).}

\bibitem{cras}  {\small P.H. Chavanis C. R. Physique {\bf 7}, 318 (2006).}


\bibitem{murray}  {\small J.D. Murray, {\it Mathematical Biology} (Springer, Berlin, 1991).}

\bibitem{ks}  {\small E. Keller, L.A. Segel J. theor. Biol. {\bf 26}, 399 (1970).}

\bibitem{horstmann}  {\small D. Horstmann, Jahresberichte der DMV  {\bf 106}, 51 (2004).}

\bibitem{crs}  {\small P.H. Chavanis, C. Rosier and C. Sire, Phys. Rev. E {\bf  66}, 036105 (2002).}

\bibitem{crrs}  {\small P.H. Chavanis, M. Ribot, C. Rosier and C. Sire,  Banach Center Publ.  {\bf  66}, 103 (2004).}

\bibitem{mass}  {\small P.H. Chavanis, Physica A [{\tt arXiv:0706.3603}].}

\bibitem{sc}  {\small C. Sire and P.H. Chavanis, Phys. Rev. E {\bf 66}, 046133 (2002).}

\bibitem{post}  {\small C. Sire and P.H. Chavanis, Phys. Rev. E {\bf 69}, 066109 (2004).}

\bibitem{gen}  {\small P.H. Chavanis, Phys. Rev. E {\bf 68}, 036108 (2003).}

\bibitem{next}  {\small P.H. Chavanis, Physica A {\bf 340}, 57 (2004).}

\bibitem{cban}  {\small P.-H.~Chavanis, Banach Center Publ. {\bf 66}, 79 (2004).}

\bibitem{degrad}  {\small P.H. Chavanis, Eur. Phys. J. B {\bf 54}, 525 (2006).}

\bibitem{gamba}  {\small A. Gamba, D.  Ambrosi, A. Coniglio, A. de Candia, S. di Talia, E. Giraudo, G. Serini, L. Preziosi, F.A. Bussolino, Phys. Rev. Lett. {\bf  90}, 118101 (2003).  }

\bibitem{vergassola}  {\small M. Vergassola, B. Dubrulle, U. Frisch and A. Noullez, Astron. Astrophys. {\bf 289}, 325 (1994).}

\bibitem{lang}  {\small P.H. Chavanis and C. Sire, Phys. Rev. E {\bf 69}, 016116 (2004).} 

\bibitem{logotropes}  {\small P.H. Chavanis and C. Sire, Physica A {\bf 375}, 140 (2007).}

\bibitem{frank}  {\small T.D. Frank, {\it Non Linear Fokker-Planck Equations} (Springer, Berlin, 2005).}

\bibitem{epjbjeans}  {\small P.-H.~Chavanis, Eur. Phys. J. B {\bf 52}, 433 (2006).}

\bibitem{virial}  {\small P.H. Chavanis and C. Sire, Phys. Rev. E {\bf 73}, 066103 (2006); Phys. Rev. E {\bf 73}, 066104 (2006).  }

\bibitem{revue}  {\small P.-H.~Chavanis,  Int J. Mod. Phys. B {\bf 20}, 3113 (2006).}



\bibitem{schweitzer}  {\small F. Schweitzer and L.Schimansky-Geier, Physica A  {\bf 206}, 359 (1994).}

\bibitem{stevens}  {\small A. Stevens, SIAM J. Appl. Math.    {\bf 61}, 183 
(2000).}

\bibitem{grima}  {\small T.J. Newman and R. Grima, Phys. Rev. E {\bf 70}, 051916 (2006).}

\bibitem{hb}  {\small P.H. Chavanis, Physica A {\bf 361}, 55 (2006); Physica A {\bf 361}, 81 (2006).}



\bibitem{perthame}  {\small B. Perthame, Appl. Math. {\bf 49}, 539 (2004).}

\bibitem{borland}  {\small L. Borland, Phys. Rev. E {\bf 57}, 6634 (1998).}

\bibitem{tsallis}  {\small C. Tsallis, J. Stat. Phys. {\bf 52}, 479 (1988).}

\bibitem{kaniadakis}  {\small G. Kaniadakis, Physica A {\bf 296}, 405 (2001).}

\bibitem{pp}  {\small A.R. Plastino and A. Plastino, Physica A {\bf 222}, 347 (1995).}



\bibitem{csr}  {\small P.H. Chavanis, J. Sommeria  and R. Robert, Astrophys. J. {\bf 471}, 385 (1996).} 



\bibitem{lemou}  {\small P.-H.~Chavanis, P. Lauren\c cot \& M. Lemou,  Physica A \textbf{341}, 145 (2004).}

\bibitem{interpretation}  {\small P.H. Chavanis and C. Sire, Physica A {\bf 356}, 419 (2005).}

\bibitem{masscrit}  {\small P.H. Chavanis and C. Sire, [{\tt arXiv:0705.4366}]}

\bibitem{peebles}  {\small J. Peebles, {\it Large-Scale Structures of the Universe} (Princeton University Press, 1980).}

\bibitem{filbet}  {\small F. Filbet, P. Lauren\c cot and B. Perthame, J. Math. Biol. {\bf 50}, 189 (2005).}



\bibitem{pitaevskii}  {\small E.M. Lifshitz, L.P. Pitaevskii, {\it Physical Kinetics} (Pergamon Press, Oxford, 1981)}

\bibitem{jl}  {\small W. J\"ager and S. Luckhaus, Trans. Am. Math. Soc. {\bf 329}, 819 (1992).}

\bibitem{bt}  {\small J. Binney and S. Tremaine, {\it Galactic Dynamics} (Princeton Series in Astrophysics, 1987).}

\bibitem{hillen}  {\small T. Hillen and K. Painter, Adv. Appl. Math.  {\bf 26}, 280 (2001).}


\bibitem{bray}  {\small A. Bray, Adv. Phys. {\bf 43}, 357 (1994).}






\end{thebibliography}
\end{document}